\documentclass[a4paper,10pt]{iopart}
\usepackage[utf8]{inputenc}
\usepackage{color}
\usepackage{graphicx}
\usepackage{amssymb}
\usepackage[bookmarks, bookmarksopen, bookmarksnumbered, colorlinks]{hyperref}

\newcommand{\dd}{\ensuremath{\textrm{d}}}

\newcommand{\beq}{\begin{equation}}
\newcommand{\eeq}{\end{equation}}
\newcommand{\bea}{\begin{eqnarray}}
\newcommand{\eea}{\end{eqnarray}}
\newcommand{\bit}{\begin{itemize}}
\newcommand{\eit}{\end{itemize}}
\newcommand{\bfi}{\begin{figure}}
\newcommand{\efi}{\end{figure}}
\newcommand{\bfic}{\begin{figure*}}
\newcommand{\efic}{\end{figure*}}
\newcommand{\bce}{\begin{center}}
\newcommand{\ece}{\end{center}}
\newcommand{\bt}{\begin{table}}
\newcommand{\et}{\end{table}}
\newcommand{\btb}{\begin{tabular}}
\newcommand{\etb}{\end{tabular}}

\newcommand{\Mean}[1]{\ensuremath{\left\langle #1 \right\rangle}}

\newcommand{\Mtot}{\ensuremath{M_\textrm{tot}}}

\newcommand{\MADM}{\ensuremath{M_\textrm{ADM}}}
\newcommand{\vol}{\ensuremath{\textrm{vol}}}
\newcommand{\Rr}{\ensuremath{\textbf{R}}}

\newcommand{\calR}{\ensuremath{{\cal R}}}

\newenvironment{remark}[1][Remark]{\begin{trivlist}
\item[\hskip \labelsep {\bfseries #1}]}{\end{trivlist}}

\newcommand{\qed}{\nobreak \ifvmode \relax \else
      \ifdim\lastskip<1.5em \hskip-\lastskip
      \hskip1.5em plus0em minus0.5em \fi \nobreak
      \vrule height0.75em width0.5em depth0.25em\fi}

\begin{document}

\title{Nonlinear effects of general relativity from multiscale structure}

\author{Mikołaj Korzyński}

\address{
Center for Theoretical Physics, Polish Academy of Sciences \\
Al. Lotnik\'ow 32/46 \\
02-668 Warsaw \\
Poland}
\ead{korzynski@cft.edu.pl}

\begin{abstract}
When do the nonlinear effects of general relativity matter in astrophysical situations? They are obviously relevant for very compact sources of the gravitational field, such as
 neutron stars or  black holes. In this paper I discuss another, less obvious situation, in which large relativistic effects may arise due to a complicated, multiscale structure 
of the matter distribution. I present an exact solution with an inhomogeneous energy density distribution in the form of a hierarchy of nested voids and overdensities of various sizes, 
extending from the homogeneity scale down to arbitrary small scales. I show that although each of the voids and overdensities seems to be very weakly relativistic, and thus easy to
describe using the linearized general relativity, the solution taken as a whole lies in fact in the nonlinear regime. Its nonlinear properties are most easily seen when we compare
the ADM mass of the solution
and the integral of the local mass density: the difference between them, i.e. the relativistic mass deficit, can be significant  
provided that the inhomogeneities 
extend to sufficiently small scales. 
The non-additivity of masses implies a large backreaction effect, i.e. significant discrepancy between the averaged, large-scale  effective stress-energy tensor and the 
naive average of the local  energy density. 
 I show that this is a general relativistic effect
arising due to the inhomogeneous, multiscale structure. 
I also discuss the relevance of the results in cosmology and relativistic astrophysics.
\end{abstract}

\maketitle
\section{Introduction}

When do the non--linear, relativistic corrections to the Newtonian laws of gravity matter? The obvious answer is that they play an important role if
the gravitational fields in question are strong. The strength of the gravitational field in turn is determined by the distribution of matter. 
For a localized object the compactness parameter $\varepsilon$, defined as the ratio of its mass expressed in the geometric units and its physical size, serves typically as
a measure of the strength of the gravitational field. Indeed, the deviation of the metric tensor from the flat one in appropriate coordinates is typically
proportional to $\varepsilon$ and thus all relativistic corrections are of the same order or higher.

The nonlinearity of GR has many consequences. One of them is the difference between the total mass of an object measured far away from it and the sum of masses of its
constituents, by which we mean the sum of the masses of compact, discrete sources or the  integral of the mass density for continuous ones. In the Newtonian gravity,
described
by the Poisson equation, these 
two quantities are necessary equal due to the Gauss theorem. The same holds in the linearized GR, equivalent to the Poisson equation in the absence of relativistic motions. 
This picture changes slightly in the next order of perturbative expansion: the faraway mass 
turns out to be smaller that the sum of masses and the difference in the lowest order is equal to the Newtonian binding energy. This mass deficit is usually very small in comparison to
the total mass of the object and thus negligible. Obviously if a given distribution of matter can be reasonably described using linearized GR then the difference between
both masses must be small. Conversely, if the mass deficit is large then obviously the solution cannot lie in the regime of applicability of the first order approximations.

Drawing the precise boundary between the linear and nonlinear regime in general relativity is more problematic in cosmology, where we are not dealing with isolated sources,
but rather with a matter distribution which is homogeneous at very large scale, but rather complicated and inhomogeneous below that.
More precisely, we observe voids and localized overdensities of various sizes and large density contrasts. It is not a priori clear if simple criteria based on the compactness of structure present in the solution can be applied in this case. The problem has recently become a subject of controversy in view of the so--called backreaction problem in cosmology 
\cite{Ellis:2011, Green:2014aga}.  Recall that in 
modern cosmological paradigm we assume the existence of an idealized, large scale averaged metric $g^{(0)}$, representing the physical metric $g_\textrm{phys}$ with
all small ``ripples'' removed. $g^{(0)}$ is further assumed to be
of the FLRW class, i.e. perfectly isotropic and homogeneous. The Einstein equations for $g^{(0)}$ (the effective or large scale equations)
read 
\bea
 G_{\mu\nu}\left[g^{(0)}\right] = 8\pi \left( T_{\mu\nu}^{(0)} + t_{\mu\nu}\right) \nonumber
 \eea
 where $T_{\mu\nu}^{(0)}$ is the appropriate average of the local, physical stress energy tensor and the backreaction term $t_{\mu\nu}$ represents
 all corrections we need to include due to the nonlinearity of GR. Since the $T_{00}$ component is the local energy density,  its average over a region of
 cosmological scale can be reasonably defined as the ratio between the total mass, i.e. the integral of the physical $T_{00}$ with the physical volume form, 
 and the volume of the region as measured by $g^{(0)}$. On the other hand, as we will see further in this paper, 
 the properties of the background FLRW solution appearing on the left hand side are related to the properties 
 of the metric tensor  away from the inhomogeneities. This way the difference between the total mass of an inhomogeneity and its quasi--local mass measured far away is related to the difference between the average energy density,
 defined as the ratio between the total mass and volume, and the effective energy density inferred from the properties of $g^{(0)}$.

An important point about the distribution of matter in our Universe is that the structure is hierarchical: large voids are separated by walls made of smaller filaments, composed
of large galaxy clusters made themselves out of galaxies etc. Most of this structure seems to lie within the nonrelativistic, linear regime due to 
their relatively big size compared to their mass and their slow, nonrelativistic motions. One could draw from that fact the conclusion that 
the nonlinear GR corrections are negligible when dealing with 
this structure and the first or at most second order of perturbation is perfectly enough to describe it. In this paper I would like to point out that this way of
thinking can in fact be quite misleading: nested, hierarchical distribution of matter may give rise to significant amplification of the nonlinear effects of GR
even though at all scales
the structure seems very ``Newtonian'' in the sense described above. Na\"{i}ve estimates based on the small compactness parameter 
of the structure present may simply not work. Moreover, I would like to argue that even if the nonlinear corrections are not large 
estimating accurately their value requires a more subtle approach, 
combining gradual coarse--graining of the structure over large scales with a perturbative approximation scheme. 

In order to illustrate these points I present  an exact solution of the Einstein equations describing a spherical object made of dust whose distribution is not homogeneous but
rather has the form of a complicated pattern of spherical voids and overdensities. The spatial matter distribution is organized in a nested, hierarchical structure 
extending over many different scales. The solution,
which I will call the multiscale foam, belongs to the well-known Swiss-cheese class and exhibits a
curious feature of self--similarity: the overdensities present there form perfect scaled--down copies of the full, spherical object with their own 
pattern of voids and overdensities present. 
I show by explicit calculation that the relativistic mass deficit in this kind of solution may be very large even though each of the voids and overdensities, 
when considered in separation from the rest, lies within the Newtonian regime
as suggested by its very small compactness parameter. Indeed, each of the void/overdensity pairs has a very small contribution to the net mass discrepancy, 
yet it is the sum of contributions of the voids and overdensities from all levels of the nested structure which produces a large effect. 

A simple way to understand how the amplification of mass deficit arises in solutions with nested, multiscale structure  is to 
think in terms of gradual coarse--graining of the inhomogeneities. Let $g_\textrm{phys}$ denote the physical metric with inhomogeneities of all possible sizes down to
the smallest inhomogeneities scale $R_\textrm{min}$. In the next step we wipe out all inhomogeneities of size from $R_\textrm{min}$ up to  
$L_1 = R_\textrm{min}(1 + x)$, $x > 0$ being a small number of the order of 1, making sure that the parameters describing the large scale structure of the solution, like the $ADM$ mass, do not change in the process. This way we  replace the original $g_\textrm{phys}$ 
with $g_{L_1}$. By doing so we decrease the total mass by a small number, related to
the average compactness parameter of the structure at the scale of $R_\textrm{min}$ and at thus decrease a bit the mass deficit. At next step of coarse--graining we smoothen out the inhomogeneities
from $L_1$ to $L_2 = L_1 (1 + x)$, decreasing the mass deficit by another small number, see Figure \ref{fig:coarsegraining}. We repeat these steps until we hit the homogeneity
scale $R_\textrm{hom}$ after which there is no more structure to coarse--grain. At that stage the mass deficit should be very small. However,  if the number of steps needed to go from
$R_\textrm{min}$ up to the homogeneity scale $R_\textrm{hom}$ is large enough even small contributions to the mass deficit from each scale can add up to a significant net result. 
\bfi
\bce
\includegraphics[width=0.45\textwidth]{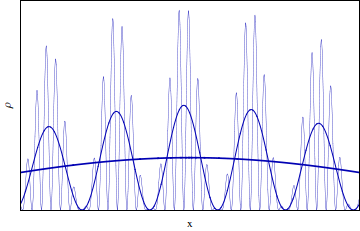} 
\caption{Two subsequent steps of the gradual coarse--graining of the inhomogeneities of the energy distribution of the solution: from the physical distribution
with inhomogeneities of all scales (thin line) to the distribution with smallest scale ripples averaged out (medium line) up to the distribution averaged up to 
the largest scales (thick line).} \label{fig:coarsegraining}
\ece
\efi

The paper is organized as follows: in the next section I review the mass deficit effect in general relativity both analytically and in the perturbative
expansion in the compactness parameter. In Section 3 I present the construction of the multiscale foam solution and discuss its properties, including the derivation and 
discussion of the mass deficit. In Section 4 I present how the perturbative expansion
can be combined with the coarse--graining approach to obtain a reasonable approximation of the mass deficit and present the numerical results showing that this kind of 
approximation gives quite accurate value of the mass deficit. 
In Section 5 I discuss the relevance of the results for cosmology and comment how they relate to the results from the papers by Green and Wald \cite{Green:2010qy} and
Ishibashi and Wald \cite{IshibashiWald}. Finally in Section 6 I summarize the conclusions of the paper. Some of the derivations
have been included in the Appendix.

In this paper will work in geometric units in which $G=1$ and $c=1$. I will assume the cosmological constant $\Lambda$ to vanish, but the general conclusions of the paper
should hold even when it is present.

\section{Non--additivity of mass in GR} \label{sec:nonadd}

 Consider time--symmetric initial data for Einstein equations with conformal ansatz
 \bea
 K_{ij} &=& 0 \nonumber\\
 q_{ij} &=& \phi^4\,\delta_{ij} \nonumber,
 \eea
 where $\delta_{ij}$ denotes the flat metric. The vacuum vector constraint equations are satisfied automatically
 \bea
 D_i\,K^{ij} = 0 \nonumber
 \eea
 while the Hamiltonian constraint reads
 \bea
 \Delta \phi = -2\pi \rho\,\phi^5 \label{eq:Hconstraint}
 \eea
 where $\Delta$ is the Laplace operator taken with respect to metric $\delta_{ij}$.
 We assume the standard fall--off condition for a compact body
 \bea
 \phi = 1 - \frac{M_\textrm{ADM}}{2r} + O(r^{-2}), \nonumber
 \eea where $\MADM$ is the ADM mass of the solution. 
 We may  integrate (\ref{eq:Hconstraint}) over a large  ball $B$, obtaining
 \bea
 \int_B \Delta \phi \,\dd^3 x &=& -2\pi \int_B \rho\,\phi^5 \,\dd^3 x. \nonumber
 \eea
 The left hand side can be reduced to a surface term which for  sufficiently large radius approaches $-2\pi  M_\textrm{ADM}$, so
 we obtain an expression for the $ADM$ mass 
 \bea
 M_\textrm{ADM} = \int \rho \,\phi^5 \,\dd^3x = \int \rho\,\phi^{-1}\,\vol \label{eq:MADMrho}
 \eea
 with the integral taken over the whole $\Rr^3$ and $\vol$ being the physical volume form given by $q_{ij}$.
 Comparing it to the definition of the total mass
 \bea
 M_\textrm{tot}  = \int \rho\, \phi^6 \,\dd^3x = \int \rho\,\vol \label{eq:Mtotrho}
 \eea
 we see that the two are in general not equal and that the difference is large when the value of
 the conformal factor deviates very much from $1$. In order to make a more direct comparison we need to adapt a perturbative approach.
 
 Consider a solution containing an object of size $R$  and ADM mass $M$. We introduce dimensionless variables $\widetilde \rho = \frac{R^3}{M}\rho$ and $\widetilde x^i = \frac{x^i}{R}$, $x^i$ being the Cartesian coordinates.
 Equation (\ref{eq:Hconstraint}) now reads
 \bea
 \widetilde\Delta \phi = -2\pi \varepsilon \widetilde\rho\,\phi^5 \label{eq:Deltaphi}
 \eea
 where $\widetilde \Delta$ is the Laplacian with respect to $\widetilde x^i$ and $\varepsilon$ is a dimensionless number called the \emph{compactness parameter}
 \bea
 \varepsilon = \frac{M}{R}. \label{eq:varepsilon}
 \eea
 As usual in general relativity, it measures the strength of the gravitational field and therefore also the scale of relativistic corrections to Newtonian gravity. 
 Indeed, if we make ansatz $\phi = 1 - \frac{\zeta}{2}\varepsilon + O(\varepsilon^2)$, we obtain in the leading order the Poisson equation for the gravitational potential $\zeta$:
 \bea
 \widetilde \Delta \zeta = {4\pi} \widetilde\rho \label{eq:Poisson}
 \eea
 or
 \bea
 \Delta (\varepsilon\zeta) = 4\pi \rho \label{eq:Poisson2}
 \eea
 in the standard, dimensional variables and coordinates. The Poisson equation does not allow any difference between the total mass and the ADM mass far away due
 to the Gauss theorem and therefore in the leading order both masses are equal. We may however insert the expansion in $\varepsilon$ for $\phi$  to (\ref{eq:MADMrho}) and (\ref{eq:Mtotrho}) and
 obtain
 this way the sub-leading terms in both masses, linear in $\varepsilon$. 
 The mass deficit expressed in dimensionless variables  becomes thus
 \bea
 \widetilde M_\textrm{tot} - {\widetilde M}_\textrm{ADM} = \int \rho \phi^5 (\phi - 1) \dd^3 \widetilde x = -\frac{\varepsilon}{2} \int \zeta\widetilde\rho\, \dd^3 \widetilde x + O(\varepsilon^2).
 \label{eq:massdeficit}\eea
 Note that the leading order contribution is of the order of $\varepsilon$ and has the form of the Newtonian binding energy of the solution of (\ref{eq:Poisson2})
 \cite{0004-637X-614-2-914}. Obviously the limit of $\varepsilon \to 0$ corresponds to switching off the nonlinear interactions in (\ref{eq:Deltaphi}). In this limit 
 the mass measured far away is exactly equal to the total mass. 
Thus the non--additivity of masses can be given a simple physical interpretation:
 due to the self--interaction of the gravitational field in general relativity, the fields in the faraway zone are sensitive not only to the matter content of the space, 
 but also to
 the energy of the gravitational field itself. In particular, in the sub-leading order the mass measured far away consists of the sum of the masses of the constituents 
 plus the negative energy of the gravitational field created by the body, i.e. minus the binding energy. This interpretation is entirely consistent with the purely geometric one
  suggested by equations (\ref{eq:MADMrho}--\ref{eq:Mtotrho}): the ADM mass is equal to the integral of the energy density with the volume form, but with a weight function
 given by the inverse of the conformal factor. Since the conformal factor is roughly 1 minus the negative Newtonian potential, the matter deep inside the potential wells 
 weighs effectively less then it would weigh far away. This causes the deficit of $\MADM$ in comparison with $M_\textrm{tot}$.

\subsection{Example -- constant density sphere}

The standard example exhibiting the relativistic mass deficit discussed in the textbooks (see \cite{waldGR, MTWch23}) is a sphere of constant density in an asymptotically
flat spacetime, described in a time--symmetric constant time slice. While it is possible to obtain the metric via the conformal ansatz above,
it is more convenient to match directly an exterior Schwarzschild solution with an interior metric of a section of a 3--sphere, i.e. a spherical cap.

\bfi
\bce
\includegraphics[width=0.45\textwidth]{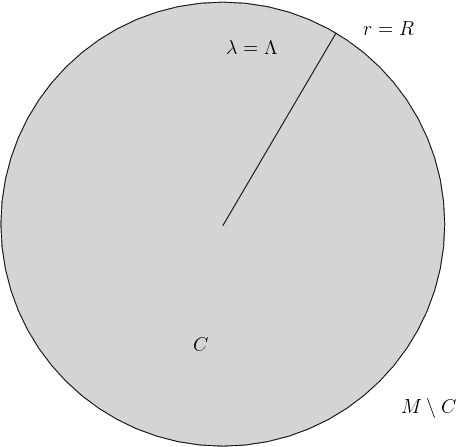} 
\caption{Matching the interior spherical cap $C$ to an exterior Schwarzschild.} \label{fig:construction1}
\ece
\efi
Let $q_M$ be the standard constant time slice of the Schwarzschild solution with mass $M$ measured in the geometric units:
\bea
q_M =  \left(1-\frac{2M}{ r}\right)^{-1}\,\dd r^2 + r^2\left(\dd \theta ^2 + \sin^2\theta\,\dd\varphi^2\right). \label{eq:qM}
\eea
 It is well known (see for example \cite{MTWch23,waldGR} that the solution outside a fixed sphere $r=R$ can be matched to an interior solution in the form of a spherical cap,
i.e. a geodesic ball excised from a round $S^3$ sphere equipped with metric
\bea
q_S =  {\calR}^2 \left( \dd \lambda^2 + \sin^2\lambda \, (\dd \theta ^2 + \sin^2\theta\,\dd\varphi^2)\right), \label{eq:qS}
\eea
$\calR$ denoting the radius of of the sphere and $(\lambda,\theta,\varphi)$ being the standard angular coordinates, see Figure \ref{fig:construction1}.
The  cap $C$ is given by $\lambda \le \Lambda$ and
 represents a spherical body of constant mass density $\rho$ determined by the value of $\calR$:
 \bea
 \rho = \frac{3}{8\pi\calR^2}, \label{eq:rhoviacalR}
 \eea
(see  \ref{sec:appendix} the derivation of all formulas in this subsection).
The matching  conditions we impose on the matching sphere $r=R$, $\lambda = \Lambda$ demand that the induced metric and its first derivative is equal
on both sides. They imply
 relations between $M$ and $R$ and the parameters $\calR$, $\rho$ and $\Lambda$ describing the interior solution, derived in  \ref{sec:appendix}. Namely,
$M$ must be related to $\rho$ via the standard formula
\bea
 M = \frac{4\pi \rho R^3}{3} \label{eq:Mviarho}
\eea
(note that since the metric $q_S$ is not flat, the expression above is \emph{not} the product of the energy density and the physical volume of $C$).
The curvature radius $\calR$  of the interior solution and the radius  $\Lambda$  measured using the angular coordinate $\lambda$ are given by
\bea
\sin\Lambda &=& \sqrt{\frac{2M}{R}}\label{eq:sinLambda}\\ 
\calR &=& \frac{R^{3/2}}{\sqrt{2M}}  = \frac{R}{\sin\Lambda}. \label{eq:calR}
\eea
Note that in the first equation we take $\Lambda$ from the branch between 0 and $\frac{\pi}{2}$.
The metric we have constructed this way will be denoted by $q_0$.

The ADM mass of the solution is equal to the Schwarzschild mass $M$, while the total mass, product of $\rho$ and the physical volume of $C$, reads
\bea
M_\textrm{tot} &=& 4\pi\calR^3\,\Xi(\Lambda)\,\rho = M \frac{3\Xi(\Lambda)}{\sin^3\Lambda}, \label{eq:Mtotball}
\eea
where
\bea
\Xi(\mu)=\frac{\mu}{2}-\frac{\sin 2\mu}{4}, \label{eq:Xidef}
\eea
the mass deficit is thus
\bea
\Delta M = M_\textrm{tot} - \MADM = M\left(\frac{3\Xi(\Lambda)}{\sin^3\Lambda} - 1 \right).\nonumber
\eea
We can expand the expressions above in the compactness parameter $\varepsilon$ to obtain
\bea
M_\textrm{tot} = M\left(1 + \frac{3}{5}\varepsilon + O(\varepsilon^2)\right). \label{eq:xexpansion0}
\eea
For the sake of convenience we also introduce the dimensionless relative mass deficit parameter
\bea
x(\varepsilon) = \frac{M_\textrm{tot} - M_\textrm{ADM}}{M_\textrm{ADM}} = \frac{3}{5}\varepsilon + O(\varepsilon^2). \nonumber
\eea
The  mass deficit is indeed of the order of $\varepsilon$, so the difference between the two masses is negligible for very weakly relativistic objects like Earth or Sun
($\varepsilon = 10^{-8}-10^{-9}$). It is 
also straightforward to verify that the first order term $\frac{3}{5}M\varepsilon = \frac{3M^2}{5R}$ is equal to
the Newtonian binding energy of a uniform ball of dust.

\section{The multiscale foam object} \label{sec:multiscale}

\subsection{Construction - first step}

\bfi
\bce
\includegraphics[width=0.45\textwidth]{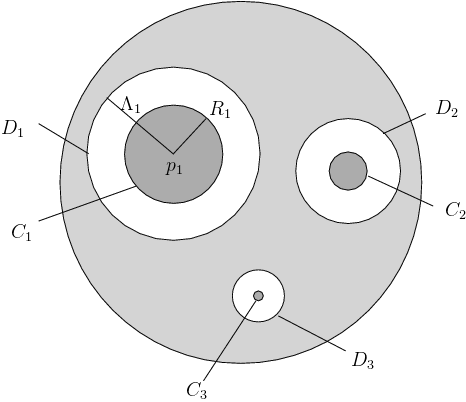} 
\caption{Next step of the construction: we replace the solution inside a cap $D_1$, centered at point $p_1$ and of radius $\Lambda_1$ measured in angular coordinates, 
by a spherical void/overdense region pair,
consisting of a shell equipped with another Schwarzschild metric matched to a spherical cap $C_1$ at area radius $R_1$. The number and positions of pairs we create this way is arbitrary
as long as they do not overlap.} \label{fig:construction2}
\ece
\efi

We begin by taking the uniform ball of dust solution $q_0$ from the previous section. In the next step we pick $n$ non--overlapping spheres (caps) $D_i$ inside the ball $C$,
centered at $n$ points $p_i$, each of radius $\Lambda_i < \Lambda$ measured in the appropriate angular coordinates centered at $p_i$, see Figure \ref{fig:construction2}. Since the metric $q_1$ inside 
$C$ is homogeneous and isotropic, the positions of the caps $D_i$ are irrelevant as long as they do not overlap. We now replace the solutions inside each $D_i$ by an 
\emph{interior} Schwarzschild solution, with the mass and the radius of matching given by
\bea
M_i &=& \frac{\calR \sin^3 \Lambda_i}{2} = M \left(\frac{\sin\Lambda_i}{\sin\Lambda}\right)^3 \label{eq:Mi}\\
\tilde R_i &=& \calR \sin \Lambda_i = R \left(\frac{\sin\Lambda_i}{\sin\Lambda}\right) \label{eq:tildeRi}
\eea
(see again  \ref{sec:appendix}). Finally we excise from the Schwarzschild metric the ball of area radius $ R_i <\tilde  R_i$ and replace it again by an $S^3$ cap $C_i$. 
In principle the choice of $R_i$ is arbitrary, but we will use this freedom to impose an additional condition by demanding each cap $C_i$ to be perfectly homothetic to the initial 
cap $C$, 
i.e. for their metric tensors to satisfy
\bea
q_1\Big|_{C_i}  = a_i \cdot \zeta_i^* q^C_{0} = a_i \cdot \zeta_i^* q_{0} \label{eq:selfsimilarity}
\eea
where $\zeta_i: C_i \to C$ is an appropriate diffeomorphism given by the identification of points with the same spherical coordinates $(\lambda,\theta,\varphi)$ on $C_i$ and $C$
and $a_i$ is a constant. Equation (\ref{eq:selfsimilarity}) implies a partial self-similarity of the solution and therefore we will
refer to it as the \emph{self-similarity condition}. 
It turns out that (\ref{eq:selfsimilarity}) can always be satisfied if we simply choose 
\bea
R_i = \gamma_i R, \nonumber
\eea
where
\bea
\gamma_i = \left( \frac{\sin \Lambda_i}{\sin \Lambda}\right)^3. \label{eq:gammai}
\eea
This can be seen rather easily if we realize that the necessary and sufficient condition for the homogeneous sphere $C_i$ to be similar to $C$ is 
the equality of their compactness parameters
\bea
\frac{M}{R} = \frac{M_i}{R_i}.\nonumber
\eea
This condition, taken together with (\ref{eq:Mi}), gives (\ref{eq:gammai}).
The rescaling coefficient $a_i$ from (\ref{eq:selfsimilarity}) is equal to $\gamma_i^2$.
The self-similarity condition implies that each cap $C_i$ has the same angular size $\Lambda$ as $C$, 
although it has a proportionally smaller curvature radius $\calR_{i} = \gamma_i\,\calR$, see Figure \ref{fig:construction3D}.

The solution $q_1$ has now the form of a ball of matter of constant density together with $n$ spherical voids, each containing a
concentric spherical overdense 
region inside.  The size of the overdense regions have been chose carefully to ensure that they
constitute $n$ scaled down copies of the original cap $C$.

\subsection{Nesting the structure - iteration}

We can now repeat iteratively the first step applied to each of the caps we obtained in the previous step, with both internal Schwarzschild 
solution as well as the internal cap scaled down appropriately. This step of the construction can be described in a simplified way as follows:
 we replace the round metric $q_1\Big|_{C_i}$ in each $C_i$ by an appropriately scaled down metric $q_1$ on $C$, i.e.
\bea
q_2\Big|_{C_i} = \gamma_i^2 \, \zeta_i^* q_1,\nonumber
\eea
where $\zeta_i$ was introduced in the previous subsection.
We thus obtain inside each $C_i$ $n$ smaller voids and overdense regions  denoted by $C_{i,j}$, $j=1,\dots,n$. As in the previous step, the geometry of $C_{i,j}$'s is
of course entirely homothetic (similar) to the original $C$, see again \ref{fig:construction3D}.  
On the other hand, we leave the metric tensor outside the caps $C_i$
unchanged:  let $\widetilde C$ denote the union of all caps $C_i$, then
\bea
q_2\Big|_{M\setminus \widetilde C} = q_1\Big|_{M\setminus \widetilde C}. \nonumber
\eea
Since the small overdense caps we obtain at each step are always similar to $C$, the replacement process can be repeated iteratively as many times as we want, yielding metric $q_N$ with $n^N$ small caps, all
homothetic to the initial $C$ and to each other. 

The iteration step can be summarized in a concise way as follows: let $N$ be an integer, $N \ge 1$, and let $q_N$ be the metric tensor after $N$th step. Let 
$a=(i,j,\dots,k)$ be a multi-index of length $N$, taking values from $1$ to $n$. Finally let $C_a$ denote the smallest cap in $q_N$, located inside $C_i$, $C_{(i,j)}$ etc. 
Let $\tilde C_N$ denote the union of all smallest caps $C_a$ in $q_N$. We then define the metric $q_{N+1}$ by conditions
\bea
q_{N+1}\big|_{M\setminus \tilde C_N} &=& q_{N}\big|_{M\setminus \tilde C_N} \nonumber\\
q_{N+1}\big|_{C_a} &=& \gamma_i^2\gamma_j^2\cdots \gamma_k^2\zeta_a^*\cdot q_1\big|_{C}, \nonumber
\eea
$\zeta_a: C_a \mapsto C$ being a diffeomorphism and $\gamma_i$ given by (\ref{eq:gammai}). 

\bfi
\bce
\includegraphics[width=0.45\textwidth, trim = 0 100 0 90, clip=true]{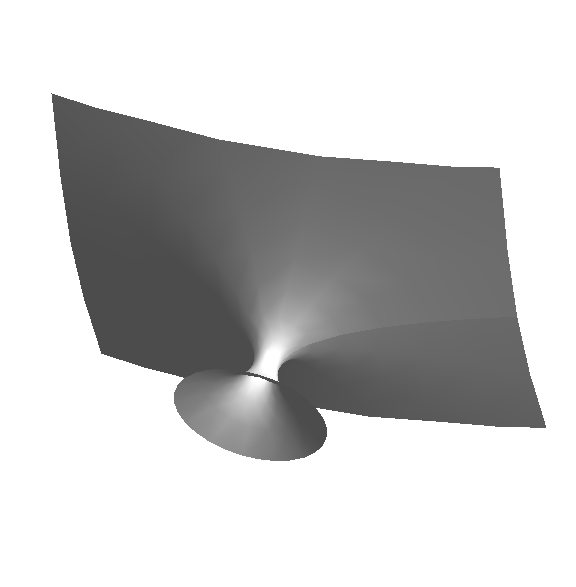}  
\includegraphics[width=0.45\textwidth, trim = 0 180 0 200, clip=true]{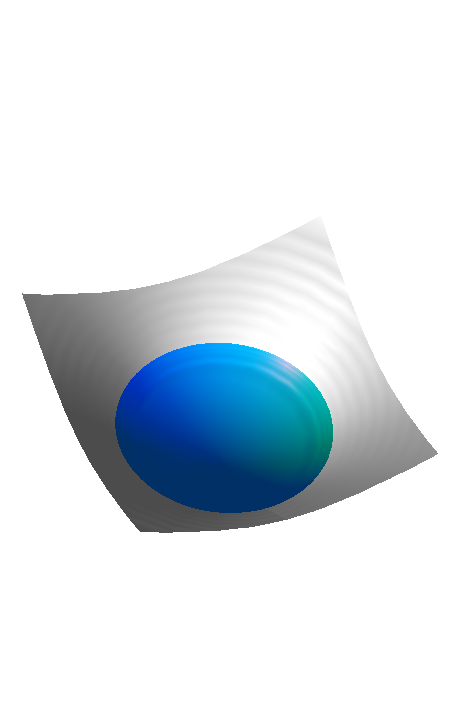} 
\includegraphics[width=0.45\textwidth, trim = 0 180 0 200, clip=true]{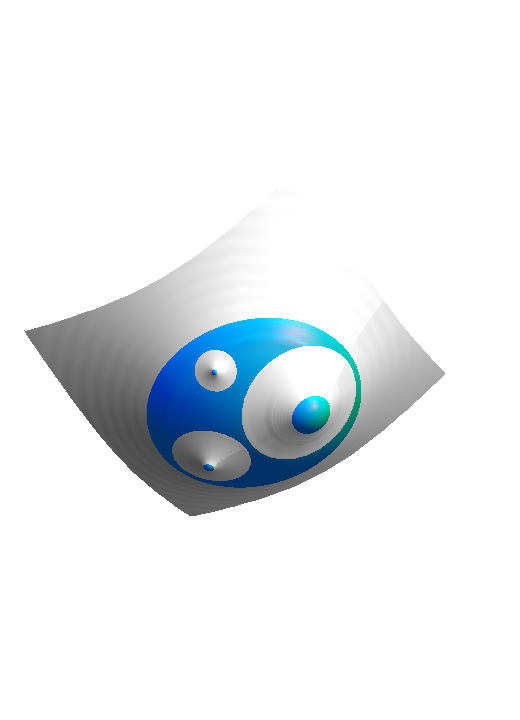} 
\includegraphics[width=0.45\textwidth, trim = 0 200 0 180, clip=true]{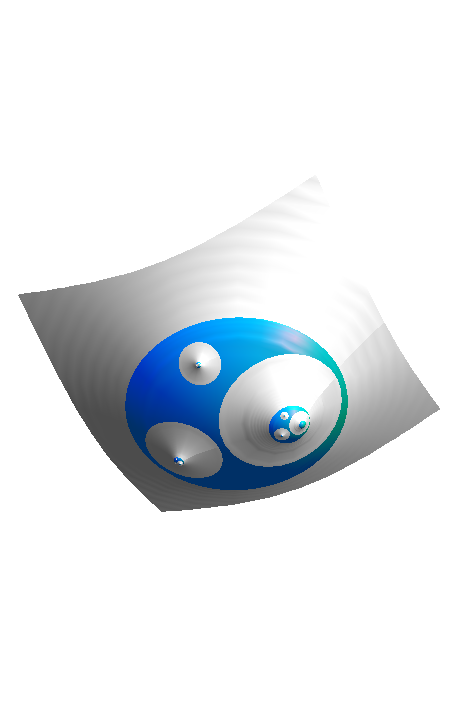} 
\caption{Construction of the multiscale foam with one dimension suppressed. The Schwarzschild solution in the form of the Flamm's paraboloid (top--left). The paraboloid
is then cut along a circle and the remaining hole is filled with an appropriately fitted blue spherical cap producing the solution $q_0$ (top--right). We then fit a number of funnel--shaped sections of 
Schwarzschild to a number disks excised from the cap. The funnels are then cut along appropriately chosen circles and the interior replaced again with matching caps (bottom--left).
The positions  of the cuts on the funnels are chosen carefully so that the small caps covered the same solid angle as the original cap, yielding $q_1$. The whole step is then 
repeated
for each of the small caps producing $q_2$. Note the self-similarity of 
the construction.} \label{fig:construction3D}
\ece
\efi


It is tempting to consider the limit $N\to\infty$, but note that the curvature of the caps grows exponentially with $N$. Therefore the limit would contain an infinite number of 
curvature singularities. We therefore consider here only finite $N$.

\subsection{Properties, time development etc.}

The solutions described above may be parameterized by the mass $M$ of the Schwarzschild solution we started from, the size of the body $R$, the angular sizes of the voids
$\Lambda_i$ created during each construction step and the nesting level of structure $N$. The compactness parameter of the initial Schwarzschild metric $q_0$ 
\bea
\varepsilon = \frac{M}{R} \nonumber
\eea
is dimensionless and may serve as a measure of the strength of the gravitational field and of the strength of the nonlinear effects of GR. Due to the
self-similarity $\varepsilon$ is actually universal throughout the whole solution: at every level of the construction we only have scaled
down void--overdense region combinations with \emph{exactly the same} compactness parameter. The situation is thus reminiscent of
classical fractals like the Sierpiński triangle or Koch curve: no matter how much and where we zoom in, all we observe is an appropriately rescaled copy
of the metric of a spherical body with the same $\varepsilon$.

The properties of the metric outside the ball $C$ remain unchanged during the construction. In particular, the ADM mass of the configuration 
is independent of the nesting level and simply equal to the mass of the initial Schwarzschild. 

The solution belongs to the Swiss cheese type of solutions, closely related to the Lema\^{i}tre--Tolman--Bondi class. The equations for its time evolution
can be solved explicitly. It turns out that the foam undergoes a complete, non--uniform collapse with the overdense regions collapsing first.

\subsection{Calculating the mass deficit} \label{sec:massdeficit}

As we pointed out before, the construction steps do not change the metric tensor outside $C$. Therefore the asymptotic properties of the metrics $q_N$ are independent of $N$. In particular, all $q_N$ have 
the same ADM mass equal to the Schwarzschild mass $M$ of the solution $q_0$. The total mass 
$M_\textrm{tot}^{(N)}$ on the other hand does change with every step of the construction and can be represented by the sum of a finite series of the form:
\bea
M_\textrm{tot}^{(N)} = M_\textrm{tot}^{(0)} + \Delta  M_\textrm{tot}^{(1)} + \cdots +  \Delta  M_\textrm{tot}^{(N)}, \label{eq:Mtotseries}
\eea
where $\Delta  M_\textrm{tot}^{(k)}$ denotes the difference of the mass deficit between $q_k$ and $q_{k-1}$. We will now demonstrate that due to the self-similarity of 
the solution in 
question the series above is a geometric series and derive an
explicit expression for the mass deficit at the nesting level $N$. 

In the first step of the construction we have removed $n$ caps $D_i$ of radius $\Lambda_i$, each containing a fraction $\alpha_i$ of the original mass $M^{(0)}_\textrm{tot}$. 
Since the matter density $\rho$ is constant in $C$ this fraction  is 
equal to the fraction of the volume of $C$ occupied by $D_i$:
\bea
\alpha_i = \frac{\vol(D_i)}{\vol(C)} = \frac{\Xi(\Lambda_i)}{\Xi(\Lambda)}, \nonumber
\eea
where
\bea
\Xi(\mu)=\frac{\mu}{2}-\frac{\sin 2\mu}{4}. \nonumber
\eea
On the other hand, all of the mass of the void/overdense region which replaced the interior of $D_i$  is contained in the overdense inner cap $C_i$.
The cap constitutes a copy of the original $C$ 
scaled down by the factor of $\gamma_i$, so its total mass is equal to the rescaled total mass of $C$. The total mass of the whole ball of dust has increased this way by
\bea
M^{(1)}_\textrm{tot}(C_i) - M^{(0)}_\textrm{tot}(D_i) = \gamma_i\,M^{(0)}_\textrm{tot} - \alpha_i\,M^{(0)}_\textrm{tot}. \nonumber
\eea
We introduce another dimensionless parameter $\delta_i$ measuring by what fraction of the initial mass the total mass has increased:
\bea
\delta_i = \frac{M^{(1)}_\textrm{tot}(C_i) - M^{(0)}_\textrm{tot}(D_i)}{M^{(0)}_\textrm{tot}} = \gamma_i - \alpha_i. \label{eq:deltaiident}
\eea
 Summing over all $D_i$'s we obtain
\bea
\Delta M^{(1)}_\textrm{tot} =  \sum_{i=1}^n \left(\gamma_i\,M^{(0)}_\textrm{tot} - \alpha_i\,M^{(0)}_\textrm{tot}\right)  = 
M^{(0)}_\textrm{tot} \sum_{i=1}^n \delta_i .\label{eq:Mtot1}
\eea
We introduce a bit of short--hand notation
\bea
\alpha &=& \sum_{i=1}^n \alpha_i \label{eq:alphadef}\\
\gamma &=& \sum_{i=1}^n \gamma_i \label{eq:gammadef}\\
\delta &=& \sum_{i=1}^n \delta_i \label{eq:deltadeff}.
\eea
From (\ref{eq:deltaiident}) we obtain an identity relating the values of the 3 resummed dimensionless parameters:
\bea
\delta &=& \gamma - \alpha. \label{eq:deltadef}
\eea
Now (\ref{eq:Mtot1}) simplifies to
\bea
\Delta M^{(1)}_\textrm{tot} =  \delta\,M^{(0)}_\textrm{tot}, \nonumber
\eea
i.e. the total mass increases by the fraction $\delta$ of the original total mass $M^{(0)}_\textrm{tot}$, equal to the total mass
a uniform ball from equation (\ref{eq:Mtotball}).

In the next steps we effectively replace the smallest caps $C_a$ by rescaled copies of $C$ from $q_1$. Due to the self--similarity of the solution 
each replacement increases the mass $M_a$ of $C_a$ by the same fraction $\delta$. Therefore
the total mass  of the solution increases by the fraction $\delta$ of
the  total mass $\tilde M_N$ of the union
of the smallest caps $\tilde C_N$:
\bea
\Delta M^{(N+1)}_\textrm{tot} =  \delta\,\tilde M_N. \label{eq:DeltaN}
\eea

Finally we derive a recursion relation for $\tilde M_N$. Let $C_a$ denote one of the smallest caps in $q_N$ and $M_a$ its total mass. The next step of the construction 
gives rise to $n$ smaller caps inside every smallest cap $C_a$, each of total mass $\gamma_i\,M_a$. Therefore the total mass contained in the smallest caps changes according 
to the formula
\bea
\tilde M_{N+1} = \sum _a M_a\,\sum_{i=1}^n \gamma_i = \gamma\,\tilde M_N. \label{eq:Mrecursion}
\eea
On the other hand it is easy to see that $\tilde M_1 = \gamma M^{(0)}_\textrm{tot}$, so
\bea
\tilde M_N = \gamma^N\,M^{(0)}_\textrm{tot}. \label{eq:tildeMN}
\eea
Taking together (\ref{eq:tildeMN}) and (\ref{eq:DeltaN}) we obtain
\bea
M_\textrm{tot}^{(N)} = M_\textrm{tot}^{(0)}\left(1 + \delta\left(1 + \gamma + \cdots + \gamma^{N-1}\right) \right), \nonumber
\eea
which can be simplified to 
\bea
 M_\textrm{tot}^{(N)} = (\MADM + \Delta M)\left(1 + \delta\,\frac{1 - \gamma^N}{1 - \gamma}\right) \label{eq:Mtotexact}
\eea
if $\gamma\neq 1$ and to 
\bea
 M_\textrm{tot}^{(N)} = (\MADM + \Delta M)\left(1 + N \delta\right) \label{eq:Mtotexact2}
\eea
otherwise.

\subsection{Calculating the volume correction}

The presence of matter affects the geometry of spacetime and thus the volume form. We will now evaluate 
the correction to the total volume of the object, including the voids, due to the inhomogeneity of matter distribution. 
We will write the total volume as a sum of a series in full analogy to the equation (\ref{eq:Mtotseries})
\bea
V_\textrm{tot} = V_0 + \Delta V^{(1)} + \Delta V^{(2)} + \cdots + \Delta V^{(N)} \label{eq:Vtotseries}
\eea
where $V_0$ is the volume of $C$ measured by $q_0$ and $\Delta V^{(k)}$ are subsequent corrections. 
Just like in the case of the total mass we can prove that the series above is a geometric series in a self--similar solution. Let us
begin by evaluating $\Delta V^{(1)}$.
After the first step the total volume consists of $C\setminus (\bigcup_i D_i)$, equipped with the initial metric $q_0$, the voids $D_i\setminus C_i$ equipped with
the Schwarzschild metric and the overdense regions
$C_i$. Since each $C_i$ is a perfect copy of $C$ scaled down by $\gamma_i$, its volume is equal to $\gamma_i^3 V_0$, so
\bea
V^{(1)}_\textrm{tot} = \left(1 - \alpha\right)V_0 + \sum_i \gamma_i^3 V_0 + \sum_i V^\textrm{Schw}_i \nonumber
\eea
or
\bea
\Delta V^{(1)} = V^{(1)}_\textrm{tot} - V_0 = \left(\sum_i \gamma_i^3 - \alpha + \sum_i \frac{V^\textrm{Schw}_i}{V_0}\right) V_0. \label{eq:Vtot1}
\eea
We introduce a short-hand notation
\bea
\nu &=& \sum_i \gamma_i^3 \label{eq:nu}\\ 
\mu_i &=& \frac{V^\textrm{Schw}_i}{V_0} \nonumber\\
\mu &=& \sum_i \mu_i \nonumber\\
\kappa &=& \nu - \alpha + \mu \label{eq:kappa}
\eea
and rewrite (\ref{eq:Vtot1}) as
\bea
\Delta V^{(1)} = \kappa V_0. \label{eq:VDeltakappa}
\eea
Repeating the same reasoning as in the previous subsection we prove that $\Delta V^{(N+1)} = \kappa \widetilde V_N$, where $\widetilde V_N$ is again the 
total volume of all smallest caps $C_a$ at step $N$. Since after each construction step every smallest cap is replaced by $n$ scaled down caps, $\widetilde V_N$ 
satisfies  a recursion relation 
\bea
\widetilde V_N = \nu \widetilde V_{N-1} \nonumber
\eea
analogous to (\ref{eq:Mrecursion}). Since $\widetilde V_1 = \kappa V_0$, we obtain the final result
\bea
V_\textrm{tot}^{(N)} = V_0\left(1 + \kappa \left(1 + \nu  + \cdots + \nu^{N-1}\right)\right) = V_0\left(1 + \kappa\frac{1-\nu^N}{1-\nu}\right). \label{eq:Vtotexact}
\eea

\subsection{Large effect from small $\varepsilon$} \label{sec:smallepsilon}

Let us consider the limit of $\varepsilon \ll 1$. For bodies without a nested structure, like a uniform sphere from Section 1, this corresponds to the weakly relativistic limit in which nonlinear effects of GR 
can be safely neglected. 

It follows from (\ref{eq:sinLambda}) that
\bea
\Lambda = \sqrt{2\varepsilon} + O(\varepsilon). \nonumber
\eea
We need to expand $\delta$ and $\alpha$ in terms of $\varepsilon$, keeping the sizes and the positions of the voids/overdensities fixed.
Let $A_i$ denote the fraction $\frac{\Lambda_i}{\Lambda}$, $A_i < 1$. We expand 
\bea
\alpha_i = \frac{\Xi(A_i\sqrt{2\varepsilon})}{\Xi(\sqrt{2\varepsilon})} = A_i^3\left(1 + \frac{2\varepsilon}{5}(1-A_i^2)\right) + O(\varepsilon^2)\label{eq:alphaiexpansion}
\eea
and
\bea
\gamma_i = \frac{\sin^3(A_i\sqrt{2\varepsilon})}{\sin^3(\sqrt{2\varepsilon})} = A_i^3\left(1 + \varepsilon(1-A_i^2)\right) + O(\varepsilon^2),\label{eq:gammaiexpansion}
\eea
thus the difference between the two expressions is of the order of $\varepsilon$:
\bea
\gamma_i - \alpha_i =    \frac{3\varepsilon}{5}A_i^3(1-A_i^2) + O(\varepsilon^2) \nonumber
\eea
and consequently coefficient $\delta$ from (\ref{eq:deltadef}) is also of the order of $\varepsilon$:
\bea
\delta =    \frac{3\varepsilon}{5}\sum_{i=1}^n A_i^3(1-A_i^2) + O(\varepsilon^2) = \frac{3\varepsilon}{5}\sum_{i=1}^n \alpha_i(1-\alpha_i^{2/3}) + O(\varepsilon^2)
\label{eq:deltaexpansion}
\eea
(we have used (\ref{eq:alphaiexpansion}) in the last equation).
Note that a similar reasoning shows that the volume correction coefficient $\kappa$ is of the order of $O(\varepsilon)$.

We are now ready to show that under additional assumptions it is possible to obtain a substantial value of the relativistic mass deficit despite very small value of $\varepsilon$. 
We will assume from now on that $\varepsilon \ll 1$, but additionally that $\alpha$ is very close to 1, i.e. the voids/overdensities take up almost the whole volume of
the initial sphere $C$. We will show in the next section how this can be effectively achieved, as for now we will just assume that 
\bea
\alpha = 1 - O(\varepsilon) \nonumber
\eea
(we use the minus sign here to express the fact  that $\alpha < 1$ by definition, assuming tacitly that the $O(\varepsilon)$ part is positive).

Since $\delta = O(\varepsilon)$ and $\delta > 0$, see (\ref{eq:deltaexpansion}), there are three possibilities concerning the value of $\gamma = \alpha + \delta$:
either its value is lower then 1, i.e. $\gamma = 1 - O(\varepsilon)$,  $\gamma = 1$ or $\gamma = 1 + O(\varepsilon)$ (again we abuse the Landau notation slightly by assuming 
that the value of $O(\varepsilon)$ is always positive). We will consider all three cases separately.

\textbf{Case 1: }$\gamma < 1$. The fraction $\frac{\delta}{1 - \gamma}$ appearing in (\ref{eq:Mtotexact}) is positive and $O(1)$ when expanded in $\varepsilon$. On the other hand, the expansion of $1 - \gamma^N$ yields $N\cdot O(\varepsilon)$.
Thus, if $N$ is at least of the order of $\varepsilon^{-1}$ then the mass deficit is $O(1)$, i.e. can be significant no matter how small $\varepsilon$ is. Note also that if we allow $N$ to diverge to infinity, the total mass from (\ref{eq:Mtotexact}) approaches a finite value of  $(\MADM + \Delta M)\,\frac{\delta}{1 - \gamma}$, which is itself $O(1)$. Therefore increasing the nesting level $N$ beyond $\varepsilon^{-1}$ doesn't 
increase the value of $\Mtot$ significantly, see Figure \ref{fig:xy}

\textbf{Case 2: }$\gamma = 1.$ From (\ref{eq:Mtotexact2}) we see that $M_\textrm{tot}^{(N)}$ is of the order of $N \delta = N\cdot O(\varepsilon)$. Again, if $N \approx \varepsilon^{-1}$
or larger, we obtain a significant effect. Unlike the first case however $\Mtot$ can be made arbitrary large  by choosing appropriately
large value $N$ (Figure \ref{fig:xy}).

\textbf{Case 3: }$\gamma > 1.$ Just like in the first case we notice that $\frac{\delta}{\gamma-1}$ is $O(1)$. The remaining part, i.e. $\gamma^N - 1$ is of the order of $1$ if 
$N$ is of the order of $\varepsilon^{-1}$ or higher.  We thus obtain a configuration with large mass deficit from a very small $\varepsilon$. 
It follows from (\ref{eq:Mtotexact}) that, just like in case 2, $\Mtot$ can be made as large as we want if the structure is sufficiently 
deeply nested, see again Figure \ref{fig:xy}.

\bfi
\bce
\includegraphics[width=\textwidth]{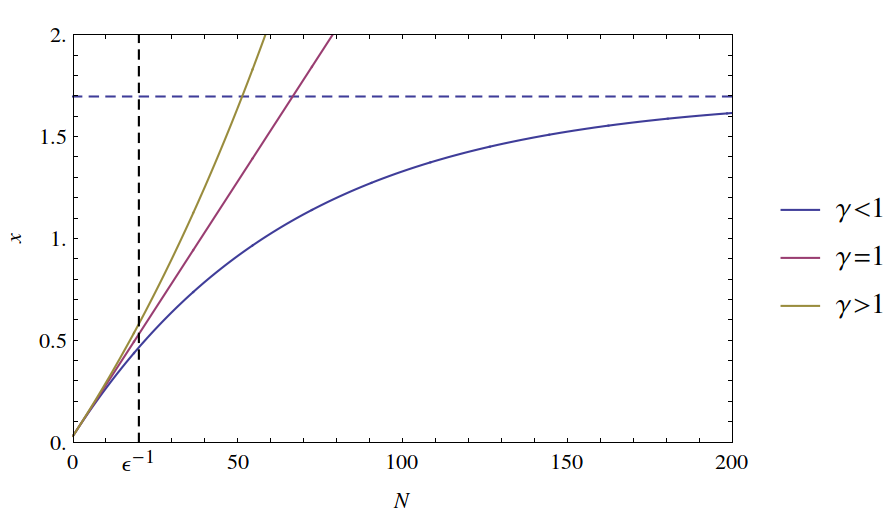} 
\includegraphics[width=\textwidth]{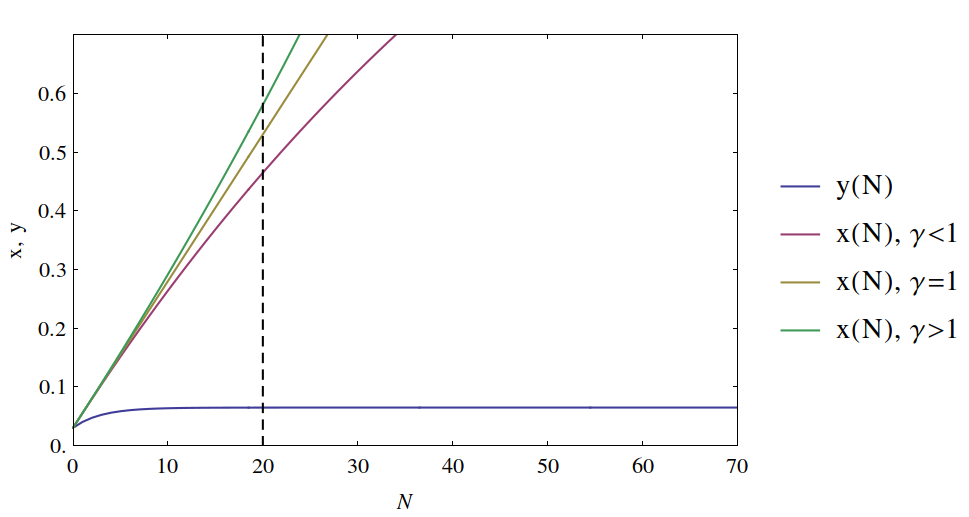} 
\caption{Upper plot: $x$ plotted as a function of $N$ for $\varepsilon = 0.05$. The mass deficit grows
from values $O(\varepsilon)$ for small $N$ up to $\approx 0.5= O(1)$ for $N \approx \varepsilon^{-1}$. For even larger $N$ 
$x$ approaches the asymptotic value of $\frac{\delta}{1-\gamma}$ if $\gamma < 1$ (dashed horizontal line) or
grows unbounded if $\gamma \ge 1$. Lower plot: Unlike the mass deficit $x$, the volume correction $y$ saturates quickly at the order
of $O(\varepsilon)$ never reaching any substantial values. Note also that both $x$ and $y$ do not vanish for $N=0$ because of the non--vanishing mass and volume discrepancies
of the order or $O(\varepsilon)$
for a uniform ball of dust.} \label{fig:xy}
\ece
\efi

It remains to show how we can obtain $\alpha$ very close to 1. Obviously this requires packing into $C$ a large number of spherical void/overdense region 
pairs as close as possible. 
The simplest construction known in mathematical literature is the well--known Apollonian sphere packing \cite{Boyd1973CJM, Boyd1973MC}. It begins with a ball containing four smaller balls which are internally tangent to it and which are pairwise externally tangent (oscullatory). In the simplest case the four balls are identical and their centers form a tetrahedron. We then 
successively add a ball of maximal radius, tangent externally to 4 already existing balls, to the configuration, see Figure \ref{fig:apollonian}. This can be done
quite effectively by the means of an appropriately chosen set of  sphere inversions \cite{ Borkovec} (see also Chapter 18 of \cite{ Mandelbrot} for two--dimensional disk packings). It is well known that in the limit of the number of balls diverging to infinity the 
total volume of the small balls approaches the volume of the initial ball, i.e. the packing asymptotically exhausts its whole volume, see Figure \ref{fig:numapollonian}.
\bfi
\bce
\includegraphics[width=0.45\textwidth]{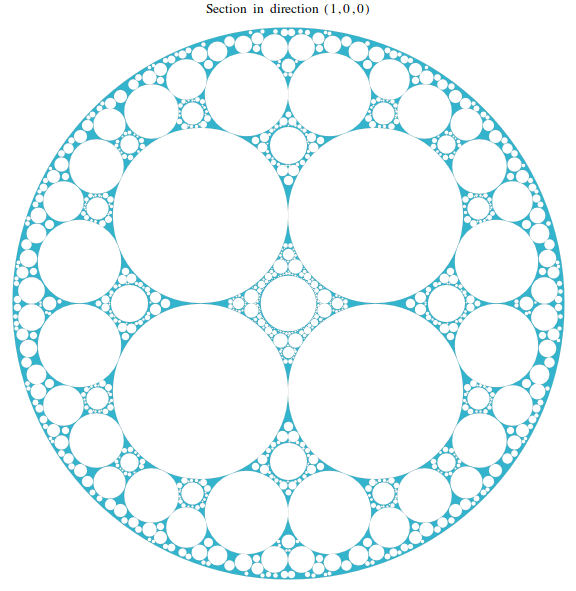} 
\includegraphics[width=0.45\textwidth]{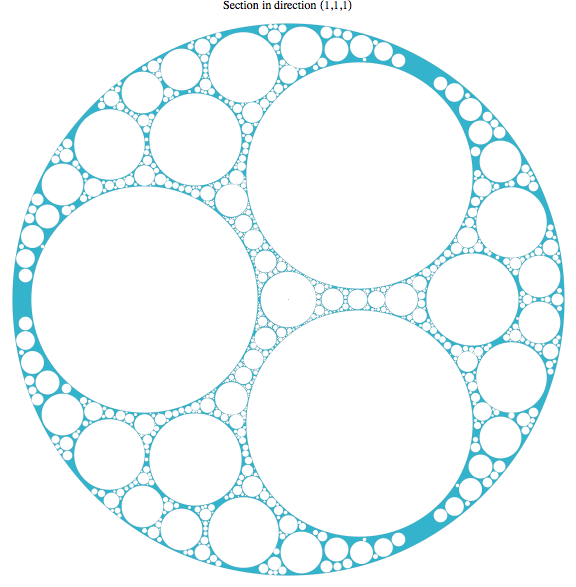} 
\includegraphics[width=0.45\textwidth]{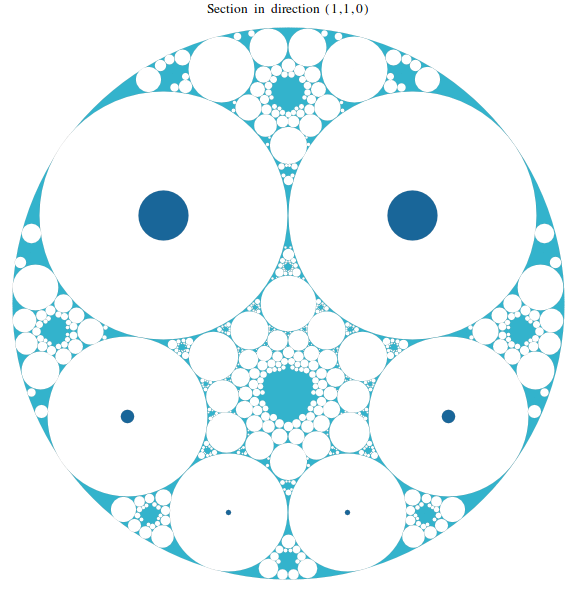} 
\includegraphics[width=0.45\textwidth]{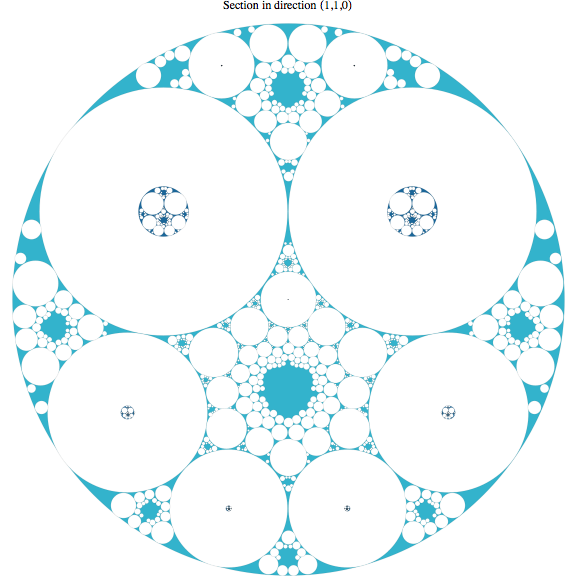} 
\caption{Upper row: two sections of the Apollonian sphere packing with 84023 spheres through the center of the large sphere. Lower row: The foam constructed from this packing with 
the nesting level $N=1$ in a different section (left) and the nesting level $N=2$ (right).}  \label{fig:numapollonian}
\ece
\efi
The classical Apollonian packing construction is performed on a ball excised from flat $\Rr^3$, equipped thus with a flat metric tensor $g$. In our case
 we
begin with a cap equipped with a constant positive curvature metric $q_0$. This is not a significant problem however, as both metrics are conformaly equivalent
via a stereographic projection which maps spheres into spheres. More precisely,
\bea
S: C \ni (\lambda,\theta,\varphi) \to (r(\lambda),\theta,\varphi)\in \Rr^3\qquad r(\lambda) =  2 \calR \tan \frac{\lambda}{2} \nonumber
\eea
is a diffeomorphism mapping $C$ into the ball $B(0,2\calR \tan{\frac{\Lambda}{2}}) \subset \Rr^3$ and satisfies $S^* g = F\,q_0$ where  $F$ is a positive function. Moreover,
any subset $W \subset B(0,2\calR \tan{\frac{\Lambda}{2}})$ is a ball iff $S(W)$ is a spherical cap in $C$. We may therefore use $S^{-1}$ to lift any set of oscullatory balls $\{B_i\}$ inside 
$B(0,2\calR \tan{\frac{\Lambda}{2}})$ to a corresponding set of oscullatory caps $\{C_i\}$ in $C$. The caps exhaust the volume of $C$, so for a sufficiently large number of them it is possible to obtain $\alpha$ as close to 1 as we need.

\bfi
\bce
\includegraphics[width=0.45\textwidth]{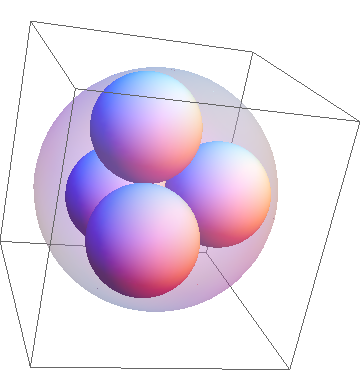} 
\includegraphics[width=0.45\textwidth]{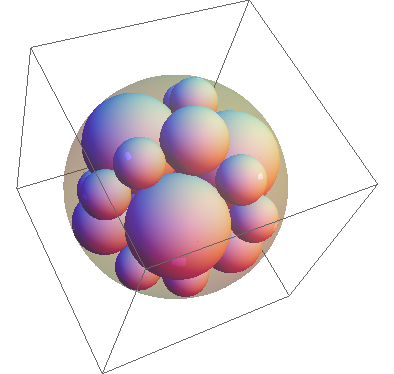} 
\caption{Apollonian sphere packing: initial 4 spheres and further steps of the construction.} \label{fig:apollonian}
\ece
\efi




\begin{remark}
One could try to make $\alpha = 1 - O(\varepsilon)$ using a simpler geometric setup by creating at each iteration step a \emph{single} void/overdensity pair $D_1$ of radius very close 
but slightly smaller than $\Lambda$, thus encompassing almost the whole of $C$. Nesting this structure would lead to a configuration of self-similar shells, see 
Figure \ref{fig:shells}. However
 direct calculations reveal that this choice \emph{does not} lead to configurations with large mass deficit for arbitrary small $\varepsilon$. The reason can be seen by inspecting
 equation (\ref{eq:deltaexpansion}): if $A_1$, which is the ratio of the angular radii of $C$ and $D_1$, is close to 1, i.e. $A_1 = 1 - O(\varepsilon) < 1$, the term 
 $1 - A_1^2$ in (\ref{eq:deltaexpansion}) is $O(\varepsilon)$ too. Consequently $\delta$ is $O(
\varepsilon^2)$ and $\gamma = \alpha + \delta = 1- O(\varepsilon) < 1$. It is straightforward to check that the total effect in this case, calculated from
equation (\ref{eq:Mtotexact}), is of the order of $O(\varepsilon)$ 
no matter how deep the nesting is. 
Quite curiously, it turns out that
the effect of large mass deficiency from nested weakly relativistic configurations requires the complicated geometry of Apollonian packing and does not work for 
simpler setups.
\end{remark}
\bfi
\bce
\includegraphics[width=0.45\textwidth]{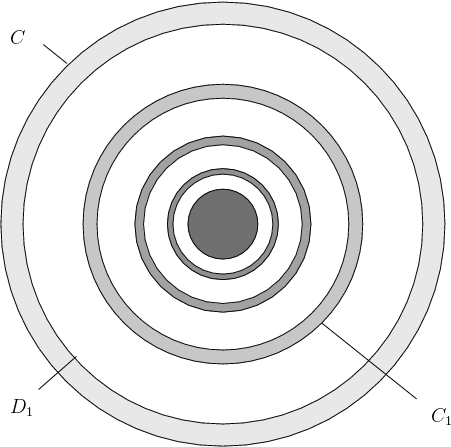} 
\caption{An attempt to obtain a large mass deficit from a nested weakly relativistic configuration in the form of self--similar concentric shells of 
increasing density.} \label{fig:shells}
\ece
\efi

Finally let us note that in contrast to the total mass the total volume corrections for the multiscale foam solution cannot grow beyond $O(\varepsilon)$. 
From (\ref{eq:nu}--\ref{eq:kappa}) we
know that 
\bea
\nu = \kappa + \alpha - \mu. \nonumber
\eea
Recall that $\kappa = O(\varepsilon)$. The sum of the first two terms is then $1 + O(\varepsilon)$, but the third term, corresponding to the volume fraction of $C$ occupied by the voids, does not vanish for 
$\varepsilon \to 0$ and thus its expansion in $\varepsilon$ reads $\mu^{(0)} + O(\varepsilon)$ with the leading order term $0 < \mu^{(0)} < 1$. Therefore  $\nu = 1 - \mu^{(0)} + O(\varepsilon)$ and for small compactness parameter it is strictly smaller than
1. Since the leading term of $1-\nu$ is of the order of $1$, no cancellation of $\varepsilon$ between $\kappa$ and $1 - \nu$ occurs in equation (\ref{eq:Vtotexact}) and $V^{(N)}_\textrm{tot}$ remains
of the order of $\varepsilon$ for any $N$, approaching the limiting value of $\frac{V_0\kappa}{1-\nu} = O(\varepsilon)$. In Figure \ref{fig:xy}
we plot the relative mass and the volume correction coefficients in terms of $N$.

\subsection{Dimensionless parameters}

Let us now have a look at the expansion of the relative mass deficit as the function of $\varepsilon$. In the leading order it reads
\bea
 x = \frac{3\varepsilon}{5} + \frac{\delta}{1 - \gamma} \left(1 - e^{N \ln \gamma}\right) + O(\varepsilon^2). \nonumber
\eea
The first term, corresponding to the binding energy of a ball of dust (see (\ref{eq:xexpansion0})), is negligible in comparison with the second one for large $N$ and we will discard it further on.
Since $\delta$ is of the order of $\varepsilon$ (see equation (\ref{eq:deltaexpansion})) and by assumption $\gamma = 1 + \gamma^{(1)}\,\varepsilon
 + O(\varepsilon^2)$, we can simplify the equation above to
\bea
 x &=& \frac{3}{5\gamma^{(1)}}\left(e^{N \varepsilon \gamma^{(1)}}-1\right) \sum_{i=1}^n \alpha_i(1-\alpha_i^{2/3})  + O(\varepsilon^2) = \nonumber\\
&=& \frac{3}{5}N \varepsilon  \sum_{i=1}^n \alpha_i(1-\alpha_i^{2/3})  + O(\varepsilon^2) \label{eq:xexpansion1}
\eea
Note that $N$ can be expressed by the ratio between the scale of the object $R$ and the scale of the smallest structure present
in the foam, given by the radius of the smallest void/overdensity pair $R_\textrm{min}$:
\bea
N = \frac{\ln {R/R_{\min}}}{\ln {R/R_n}} \nonumber
\eea
where $R_n$ is the area radius of the smallest void created at the first iteration step. 
The ratio $\ln{R/R_n}$ depends only on the geometry of the voids we create at each step. On the other hand the expression $\ln R/R_{\min}$ can be considered another dimensionless quantity after $\varepsilon$ characterizing the system. More precisely, it quantifies the distance between the homogeneity scale of the foam and the scale of its smallest ripples. We will call it the \emph{depth of structure} and denote by $D$:
\bea
D = \ln \frac{R}{R_\textrm{min}}. \label{eq:Ddef}
\eea
Thus (\ref{eq:xexpansion1}) becomes
 \bea
x(\varepsilon,D)&=& \frac{3\,\varepsilon D}{5 \ln{(R/R_n)}}  \sum_{i=1}^n \alpha_i(1-\alpha_i^{2/3})   + O(\varepsilon^2). \label{eq:xDepsilon}
\eea
The expansion in $\varepsilon$ turns out to be in fact an expansion in the product $\varepsilon D$. The analogy with quantum field theory and the renormalization group approach
suggests the interpretation of this result in terms of dressing of the coupling constants: in the presence of nested structure extending over many scales the value of the compactness parameter entering 
expressions like (\ref{eq:xexpansion0}) or (\ref{eq:xexpansion1}) becomes dressed due to nonlinear relativistic effects. It is now obvious that no matter
how small $\varepsilon$ is, we can compensate its small value by taking $D$ sufficiently large to obtain a substantial mass deficit $x$.

\subsection{In the conformal coordinates} \label{sec:conformalcoords}

It is very instructive to look at the foam solution in the conformal gauge we have discussed in Section \ref{sec:nonadd}. 
$q_N$ is indeed conformally flat and can be expressed in the form of
\bea
\left(q_N\right)_{ij} = \phi_N^4 \delta_{ij} \label{eq:conf}
\eea
with a single function $\phi_N$ encoding the whole geometry of the solution.
The conformal factor $\phi_N$ can be constructed by recursion relations in the following way.

We begin we expressing the initial homogeneous ball solution $q_0$ in conformal coordinates. Let $M$ be again its ADM mass and $R$ its area radius. We introduce
its conformal radius $\widehat R = \frac{R}{2}\left(1 - \frac{M}{R} + \sqrt{1 - \frac{2M}{R}}\right)$ and
\bea
\phi_0(r) = \left\{\begin{array}{ll} 1 + \frac{M}{2r} & \textrm{if } r \ge \widehat R \\
               \left(1 + \frac{M}{2\widehat R}\right)^{3/2} \left(1 + \frac{2Mr^2}{\widehat R^3}\right)^{-1/2}    &  \textrm{if } r < \widehat R.
               \end{array}\right. \label{eq:phi0}
\eea
Metric (\ref{eq:conf}) with conformal factor (\ref{eq:phi0}) is isometric to $q_0$ constructed by matching in Section \ref{sec:multiscale}.
In the next step we introduce voids and overdensities by multiplying the conformal factor by a function $\chi_1$ which is equal to 1 outside all the 
voids $D_i$, but not equal inside, see Figure \ref{fig:phi}. 
For each void we introduce the function
\bea
 \beta_i(r) = \left\{\begin{array}{ll}C_i\,\frac{\phi_0(g_i r)}{\phi_0(r)} & \textrm{if }r \le \widehat R_i \\
                      1 & \textrm{if } r > \widehat R_i
                     \end{array}\right., \nonumber
\eea
where $\widehat R_i$ is the radius of the void/overdensity pair in conformal coordinates and the constants $C_i$ and $g_i$ are functions 
of $M$, $\widehat R$, and $\widehat R_i$ which ensure that $\beta_i$ is continuous together with its first derivative across $r=\widehat R_i$ (this is always possible,
although the exact expressions for them are quite complicated and irrelevant here), see Figure \ref{fig:phi}.

Let now $S_i$ denote a rotation of the spherical cap $C$ which takes its center $p$ to the center $p_i$ of the overdensity $D_i$. $S_i$ is only defined
on a subset of $C$, but it can be uniquely extended to a proper conformal transformation $\Xi_i$ of the  underlying flat metric. We can now lump together 
all $\beta_i$ functions into a single one, defined as their product:
\bea
 \chi_1 = \prod_{i=1}^n \beta_i \circ\Xi_i. \nonumber
\eea
Now the conformal factor $\phi_1 = \phi_0 \chi_1$ substituted to (\ref{eq:conf}) yields a metric isometric to $q_1$.

In order to construct functions $\chi_2$ and higher, yielding the next steps of the construction, we only need to scale down functions $\beta_i$ and place them correctly.
Let $T_i$ be the rescaling
$T_i: \Rr^3 \ni x^k \mapsto g_i\,x^k \in \Rr^3$ with factor $g_i$ and let $F_i = \Xi_i\circ T_i$. $F_i$ maps $C$ into the appropriate smaller cap $C_i$. We now take
\bea
 \chi_N &=& \prod_{i_1,\dots,i_N}^n \beta_{i_1} \circ\Xi_{i_1}\circ F_{i_2}^{-1}\circ \cdots \circ F_{i_N}^{-1}\nonumber \\
 \phi_N &=& \phi_0\,\chi_1\cdots\chi_N. \nonumber
\eea

Note that each of the $\chi_k$ is itself of the order of $1 + O(\varepsilon)$ inside the overdense spheres of appropriate nesting level, so inside the \emph{smallest} 
overdense spheres the conformal factor 
is of the order of $(1 + O(\varepsilon))^N$. For a deeply nested structure ($N \approx \varepsilon^{-1}$) 
this means that in the vicinity of the
smallest overdensities the 
conformal factor deviates significantly  from $1$, see Figure \ref{fig:phi}. This means of course that the solution lies beyond the regime of validity of simple GR linearized
around a flat solution, although the total volume of space where $\phi$ deviates significantly from 1 is
rather small.
\bfi
\bce
\includegraphics[width=0.8\textwidth]{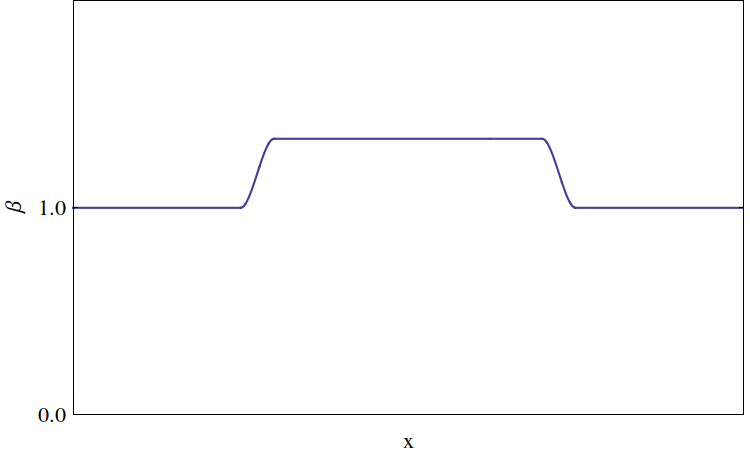}
\includegraphics[width=0.8\textwidth]{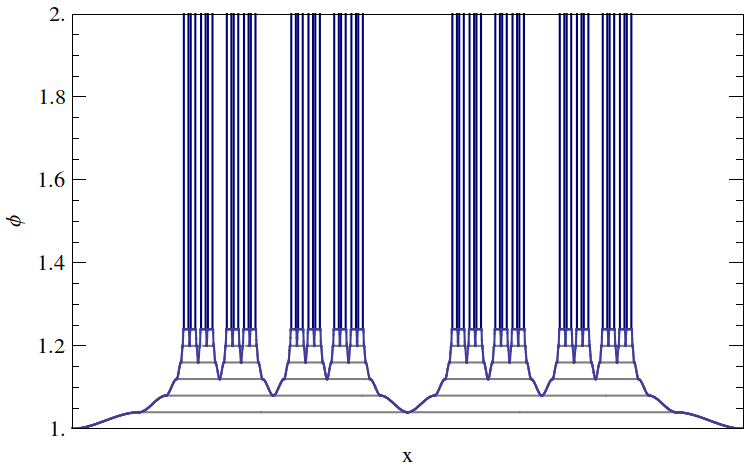} 
\caption{Upper plot: Each $\beta_i$  is a bump--like function equal to 1 outside a ball. Lower plot: The conformal factor $\phi_N$ obtained 
as a product of rescaled functions $\beta_i$. The resulting function 
only modestly deviates from $1$ in most places except the vicinity of the smallest scale overdensities, where $N$ small contributions of
the order of $O(\varepsilon)$
accumulate to a large value $O(1)$.} \label{fig:phi}
\ece
\efi

We arrive thus at the following conclusion: in the conformal gauge each single void/overdensity pair can be described by a small, quasi--Newtonian perturbation of the conformal
factor, but if
we try to extend this type of approach to the whole multiscale foam we note that it fails on the global level. 
The problem is that it is impossible to extend a conformal coordinate system up to the scale of the whole object without violating the smallness condition 
for the conformal factor. Such extension would be possible only if we somehow wiped out the finest structures in the matter distribution, i.e. adapt a coarse--graining approach
to the problem, which will be discrussed in the next section. This problem with the conformal gauge has already been noted,
see the paper by Buchert, Ellis and van Elst \cite{Buchert2009} in the memory of J\"urgen Ehlers.

\subsection{Multiscale foam in a Swiss-cheese  cosmological model}

It is possible to embed the multiscale foam in a Swiss--cheese cosmological model. Consider a closed FLRW model at the moment of its largest expansion, with the spatial metric
of the form of a 3--sphere of radius $\calR$ given by (\ref{eq:qS}). Let $\Lambda \ll 1$ be a small angle. We excise a densely packed (not necessary regular) lattice of caps of radius $\Lambda$ and
 fill them with multiscale foam balls with ADM mass $M = \frac{\calR}{2}\sin^3\Lambda$ and area radius $R = \calR \sin\Lambda$. This way we obtain a cosmological solution $r_N$
 with the metric tensor very close to the close FLRW model with dust, but with multiscale structure present throughout the universe, see Figure \ref{fig:swisscheese}.  More precisely,
 the metric $q_S$ from (\ref{eq:qS}) is an excellent approximation of $r_N$ everywhere except the
 smallest and densest overdense regions.
\bfi
\bce
\includegraphics[width=0.4\textwidth]{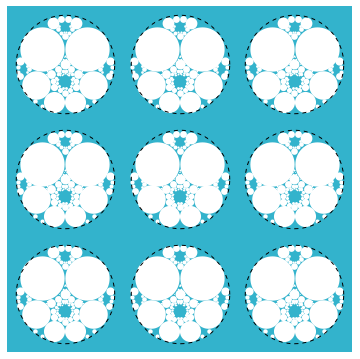}
\caption{An FLRW metric with a densely packed lattice of multiscale foam solutions.} \label{fig:swisscheese}
\ece
\efi

 Let us focus on its 
 properties at the moment of the largest expansion and consider the relation between the cosmological parameters describing the matter content inferred from 
 properties of the large scale averaged solution 
 $q_S$ and average of the local energy density. The mass of any of the excised balls calculated with the averaged metric $q_S$ is obviously $M$. On the other hand the integral of 
 the physical energy density with the physical volume form yields $M_\textrm{tot}$. As we have seen, these two values can be very different. 
 Consequently in the presence of the hierarchy of nested overdensities and voids the average energy density inferred from the averaged metric $q_S$ may differ significantly
 from the average of energy density of $r_N$. In the language of \cite{Green:2014aga} this means that a significant $00$ component of the backreaction tensor $t^{(0)}$ is
 present. This example shows that the presence of multiscale structure in cosmology may lead to accumulation of nonlinear effects of GR resulting in large backreaction terms
 in the effective FLRW equations.

\section{Coarse-graining approach to the nonlinear effects} \label{sec:coarse}

Whilst the multiscale foam solution cannot be described accurately by simple linearized gravity, it is nevertheless possible to obtain a reasonable approximation
for the mass deficit by combining the leading order perturbation in $\varepsilon$ with coarse-graining approach in which we smoothen out one by one the inhomogeneities at each 
nesting level. At each stage we need to make sure that the coarse--graining does not change the ADM mass and derive the expression for the correction of the total mass. 
Since the structure we wipe out at each stage is weakly relativistic in the sense of small compactness parameter, we may use the linear approximation of type
(\ref{eq:Deltaphi}--\ref{eq:Poisson2}) to
simplify the calculations. 

In principle this procedure should be performed in the bottom--up way: starting from the physical metric $q_N$ we coarse--grain it to $q_{N-1}$, lacking the finest overdensities
and voids, basically undoing the construction steps one by one. This way we obtain the corrections $\Delta M_\textrm{tot}^{(i)}$ from Section \ref{sec:massdeficit} in the reverse order, starting from 
$\Delta M_\textrm{tot}^{(N)}$ and ending with $\Delta M_\textrm{tot}^{(1)}$. However, as  we have noted in Section \ref{sec:massdeficit}, due to the self--similarity of the configuration 
the sum of all total mass corrections 
is given by a geometric series (\ref{eq:Mtotseries}) whose sum can be expressed via the $\alpha$, $\delta$ and $\gamma$ parameters defined by
(\ref{eq:alphadef}--\ref{eq:deltadef}). These parameters in turn can be calculated by considering only one single step of coarse--graining or construction, preferably the first one. 

\subsection{The first step}
Consider the uniform ball of dust $C$. Instead of solving the full equations we solve the first order approximation in $\varepsilon$, given by equation (\ref{eq:Poisson}).
We will work in dimensionless
variables introduced in Section \ref{sec:nonadd}, although for the sake of clarity we will omit the tildes over them. We have
\bea
\rho(x) = \left\{\begin{array}{ll}
                 \frac{3}{4\pi} & \textrm{if }\|x\| \le 1 \\
                  0 & \textrm{otherwise}
                 \end{array}\right. \nonumber
\eea
and for $\zeta$ we obtain
\bea
 \zeta(x)=\left\{\begin{array}{ll}
                  \frac{1}{2}\left(\|x\|^2 - 3\right) & \textrm{if }\|x\| \le 1 \\
                  -\|x\|^{-1} & \textrm{otherwise}
                 \end{array}\right. \nonumber
\eea
($\|\cdot\|$ denotes here the norm with respect to the Euclidean metric $\delta_{ij}$).
Let us now introduce a void/overdensity  pair centered at $p$, given by $x_p^i$ in our coordinate system. The new energy density $\widetilde\rho_a$ reads now
\bea
 \rho_a(x) = \left\{\begin{array}{ll}
                 \sigma & \textrm{if }\|x-x_p\| \le R_\textrm{in} \\
                 0 & \textrm{if }R_\textrm{in} < \|x - x_p\| \le R_\textrm{out} \\
                  \rho(x) & \textrm{otherwise.}
                 \end{array}\right. \nonumber
\eea
The conservation of the $ADM$ mass implies that
\bea
\int \rho\, \dd^3 x = \int \rho_a \,\dd^3 x\nonumber
\eea
so
\bea
\sigma = \frac{3 R_\textrm{out}^3}{4 \pi R_\textrm{in}^3} + O(\varepsilon). \nonumber
\eea
The self-similarity condition on the other hand implies that the compactness of $C$ and $C_i$ is the same:
\bea
\varepsilon \sigma R_\textrm{in}^2 = \varepsilon\rho R^2 = \frac{3\varepsilon}{4\pi} + O(\varepsilon) \nonumber
\eea
(the last equality follows from the fact that in the dimensionless variables $R=1$). 
Taken together the equations above imply that
\bea
R_\textrm{in} = R_\textrm{out}^3 + O(\varepsilon).\nonumber
\eea
The new potential $\zeta_a$ reads
\bea
\zeta_a(x) = \left\{\begin{array}{ll}
						\left(\frac{R_\textrm{out}}{R_\textrm{in}}\right)^3\frac{\| x - x_p \|^2}{2}  + F_1 &
						  \textrm{if }\| x - x_p\| \le R_\textrm{in} \\
						-R_\textrm{out}^3\,\|x - x_p\|^{-1}  + F_2 &
						  \textrm{if }R_\textrm{in} < \| x - x_p\| \le R_\textrm{out} \\ 
						  \zeta(x) &
						  \textrm{otherwise.} \\ 
						\end{array}\right. ,  \nonumber
\eea
where
\bea
F_1 &=& \frac{3}{2}\left(R_\textrm{out}^2 - 1 - \frac{R_\textrm{out}^3}{R_\textrm{in}}\right) + x_p\cdot x + \frac{1}{2}\| x_p \|^2 \nonumber \\
F_2 &=& \frac{3}{2}\left(R_\textrm{out}^2 - 1\right) + x_p\cdot x + \frac{1}{2}\| x_p \|^2\nonumber
\eea
and the scalar product is again given by the metric $\delta_{ij}$.
We can now evaluate the difference between the total mass before and after the introduction of the void: since the ADM mass
does not change, we may use (\ref{eq:massdeficit}) to obtain
\bea
M_\textrm{tot,a} - M_\textrm{tot} = -\frac{\varepsilon}{2} \int \left(\rho_a \zeta_a - \rho \zeta \right) \dd^3 x .\nonumber
\eea
Unsurprisingly the difference is simply equal to the increase of the Newtonian binding energy due to compression of the matter from the void into a ball of smaller radius.
In our case it reads
\bea
M_\textrm{tot,a} - M_\textrm{tot} = \frac{3\varepsilon}{5} R_\textrm{out}^3\left(1 - R_\textrm{out}^2\right) + O(\varepsilon^2).\nonumber
\eea
Since the volume fraction is
\bea
\alpha_i = R_\textrm{out}^{-3}\left(1 - \frac{9\varepsilon}{10}\left(1 - R_\textrm{out}^2\right)\right) + O(\varepsilon^2) \label{eq:alphalinear}
\eea
we obtain
\bea
M_\textrm{tot,a} - M_\textrm{tot} = \frac{3\varepsilon}{5}\alpha_i(1-\alpha_i^{2/3}) + O(\varepsilon^2). \nonumber
\eea
Note that in the dimensionless variables the total mass is $M_\textrm{tot} = 1 - O(\varepsilon)$, so
\bea
\delta_i = \frac{3\varepsilon}{5}\alpha_i(1-\alpha_i^{2/3}) + O(\varepsilon^2) \nonumber
\eea
and for $n$ non--overlapping void/overdensity pairs we can
express the parameter $\delta$ as
\bea
\delta = \frac{3\varepsilon}{5}\sum_{i=1}^n \alpha_i(1-\alpha_i^{2/3}) + O(\varepsilon^2). \label{eq:deltalinear}
\eea
It remains to express the rescaling coefficients $\gamma_i$ by $R_\textrm{out}$ and $\varepsilon$. This can be done by repeating the reasoning from Section \ref{sec:massdeficit}: 
we note first that that $\gamma_i$ is given by the ratio of the masses of 
$C$ before we create the void/overdensity pair $M_\textrm{tot}(C) = 1$ and the mass of $C_i$, i.e.
\bea
\gamma_i = \frac{M_\textrm{tot}(C_i)}{M_\textrm{tot}(C)} = \frac{\alpha_i\,M_\textrm{tot}(C) + \delta_i\,M_\textrm{tot}(C)}{M_\textrm{tot}(C)} = \alpha_i + \delta_i. \nonumber
\eea
This identity is true in all orders of expansion in $\varepsilon$. It follows that the summed up coefficients satisfy
\bea
\gamma = \alpha + \delta, \label{eq:gammalinear}
\eea
which, together with (\ref{eq:alphalinear})  and (\ref{eq:deltalinear}), provides a first order approximation for $\gamma$ in $\varepsilon$.
We can now repeat the whole reasoning leading to equations (\ref{eq:Mtotexact}) and (\ref{eq:Mtotexact2}) for $M_\textrm{tot}$ as a function of dimensionless $\gamma$ and $\delta$. Substituting
the Taylor expansions (\ref{eq:deltalinear}), (\ref{eq:alphalinear}) and (\ref{eq:gammalinear}) to these equations yields the approximate expression for the mass deficit.

\begin{remark}

A simpler but more crude way to obtain the mass deficit in the linear approximation would be to skip the gradual coarse--graining and apply the linearized equation (\ref{eq:Poisson}) around the vacuum solution 
to the whole  multiscale structure at once, with the right--hand side containing all the void/overdensity pairs down to the smallest ones. 
The potential $\zeta$ would contain in this case the contributions from all voids and overdensities down to the smallest scales lumped together in an additive way.
Relatively easy calculation, which we will omit here, lead to the expression (\ref{eq:xexpansion1}) for the relative mass deficit. Note that in this approach the 
approximate result is
linear also in $N$, so we do not obtain the exponential dependence of the mass deficit on the nesting level from
(\ref{eq:Mtotexact}). Consequently the mass deficit may always grow unbounded and, unlike in the coarse--graining approach, it is impossible to distinguish between the
three cases discussed in Section \ref{sec:smallepsilon}.
This demonstrates the advantage of the gradual coarse--graining approach with first order perturbations over the simple--minded first order perturbation approximation.

\end{remark}

\subsection{Numerical results}

We considered a foam with reasonably small value of $\varepsilon = 0.05$ and 84023 spherical void/overdensity pairs distributed according to the simplest Apollonian packing, 
see Figure \ref{fig:numapollonian}. The efficient algorithm for generating the Apollonian packing using 4 seed spheres and 5 generating sphere inversion transformations, working in 
  polyspherical coordinates,  was 
borrowed from Borkovec, de Paris and Peikert \cite{Borkovec}. 
The volume fraction $\alpha$ taken up by the voids is equal to 0.901683 in this solution, while  $\delta = 0.0257693$, i.e. of the order of $\varepsilon$. 
With the nesting level of $N=38$ we obtain the mass deficit fraction of $x =0.339051$, very close to $1/3$. This value is indeed significantly larger than $5\%$ suggested
by the value of the compactness. 
We also calculated the value of $x$ in the linear approximation using the coarse--graining method from Section \ref{sec:coarse} in order to
assess its accuracy, obtaining a reasonable  value of $x = 0.314274$.

\section{Relevance of the result}

\subsection{Structure in the Universe}

Do effects of this kind matter in cosmology? It is difficult to give a definite answer without a general formalism for estimating the backreaction effects and 
without the details of the microscopic distribution of matter on various scales. We may however try to get a rough order--of--magnitude estimate using 
the first order expansion formula (\ref{eq:xDepsilon}). The structure in the observable Universe is not self--similar, but as we noted
before, it is multiscale and nested. 
We can guess that (\ref{eq:xDepsilon}) remains valid in this case if we simply replace the constant compactness parameter $\varepsilon$ with a suitable average, namely
\bea
x = c\, \Mean{\varepsilon}D. \nonumber
\eea
$c$ denotes here a constant depending on the details of the geometry of the matter distribution on various scales, $\Mean{\varepsilon}$ is the average compactness 
parameter of the structure taken over a large portion of the Universe \emph{and} over all scales, while $D$ is the depth of structure defined by (\ref{eq:Ddef}). 
$D$ can be estimated as follows: the size of the largest structures observed in the Universe is around  2000Mpc \cite{Horvath:2013kwa} and we will take this number to be the end of the cosmic structure in the large scales. The end of structure in the small scales is more difficult to pin down, but we will assume here that the nested structure ends with the scale of individual stars \footnote{Note that stars are the smallest compact objects responsible for the discrete nature of matter distribution in galaxies.}. The average distance between the two neighbouring stars in the Galaxy is around 1pc, which then corresponds to the depth $D \approx 20$. While this number may seem rather unimpressive in comparison to the
compactness parameters of galaxies or stars ($10^{-5}$--$10^{-8}$), note that it may effectively boost the 
net backreaction by an order of magnitude in comparison to the most na\"{i}ve estimates basing on the value compactness.

\subsection{Relation to the Ishibashi--Wald argument}

In a reaction to a longer discussion on the issue of cosmological backreaction initiated by Buchert (see for example \cite{Buchert:2007ik, Ellis:2011} for references) Ishibashi and Wald  published a paper \cite{Ishibashi:2005sj} where they argued that the impact of inhomogeneities on the large scale dynamics of the Universe is small. Their argument 
relies on a detailed discussion on the validity and physical implications of the conformal ansatz for the perturbed FLRW metric tensor
\bea
 g = -(1 + 2\Psi) \dd t^2 + a(t)^2 (1 - 2\Psi) \gamma_{ij} \dd x^i\,\dd x^j \label{eq:WaldIshibashiAnsatz}
\eea
together with three smallness conditions on the scalar perturbation mode
\bea
 \left|\Psi\right| &\ll& 1 \quad \\
 \left|\frac{\partial\Psi}{\partial t}\right| &\ll& \frac{1}{a^2} D^i\Psi D_i\Psi\\
 (D^i\Psi D_i\Psi)^2 &\ll& (D^i D^j \Psi) D_i D_j \Psi.
 \label{eq:WaldIshibashiIneq}
\eea
They point out that this ansatz is entirely consistent with the assumption for the matter to be composed of locally inhomogeneous dust plus homogeneous fluid and that
it automatically implies that the large scale conformal factor satisfies simple Friedman equations without any significant backreaction.
The question is thus whether the physical metric tensor of the Universe satisfies globally these assumptions or not. 

Obviously conditions (\ref{eq:WaldIshibashiAnsatz}--\ref{eq:WaldIshibashiIneq}) are not satisfied in the vicinity of very  compact objects (neutron stars,
black holes etc.). One may argue that objects of this kind are rare and thus insignificant when we consider the properties of the Universe on the Hubble scale. However, as 
I pointed out in my previous paper \cite{Korzynski:2014}, if the matter in a cosmological model is contained in a large number
of evenly distributed compact sources, then the value of the mass deficit does not have to be small \emph{even if the metric tensor is very close to the smooth 
one almost everywhere}. 
The nonlinear mass 
deficit turns out to be a complicated function of the microscopic distribution of the compact sources. I proved that its value is negligible if 
all objects are spaced far apart, but it tends to increase  up to substantial values if the sources exhibit a tendency to cluster, see also the 
examples constructed by the method of images from \cite{Clifton:2014mza}.

In this paper I have shown that even in the absence of discrete sources or compact objects the value of the mass deficit may be large if we simply allow the conformal
factor to
deviate significantly from the coarse--grained one in some places, see the results of Section \ref{sec:conformalcoords}. These regions together take up a rather small fraction
of the solution's volume,
but nevertheless they may strongly influence the large--scale properties of the solution. As we have also noted, the conformal factor may attain large values simply because of 
the deep nesting level of the overdensities in the matter distribution, without the need of any compact sources or strong gravitational fields. 

More generally, the results of this paper and \cite{Korzynski:2014} show that if we allow the presence of small, localized regions of strong
gravitational fields in the solution, then conditions like (\ref{eq:WaldIshibashiAnsatz}--\ref{eq:WaldIshibashiIneq}), imposed on the metric tensor far away from those regions, are 
in general insufficient to guarantee small mass deficit (and possibly also other
backreaction effects). It is necessary to consider the fine details of the matter distribution in order to provide rigorous bounds on the cosmological backreaction. 
The gradual coarse--graining approach sketched in Section \ref{sec:coarse} seems to be a promising line of research in this direction.

\subsection{The Green-Wald formalism}

Green and Wald proposed in \cite{Green:2010qy} a framework for estimating how inhomogeneities in small scales influence  the dynamics
of the large scale metric, extending and formulating rigorously the approach of Isaacson \cite{isaacson1, isaacson2} and Burnett \cite{burnett}. Their framework is based
on a number of mathematical assumptions, including the existence of
a one--parameter family of solutions $g_{\mu\nu}(x,\lambda)$ in which the metric tends pointwise to the large--scale, average one $g_{\mu\nu}(x,0)$, while its first derivatives
are only assumed to be bounded for $\lambda \to 0$. The formalism, despite different formulation, is reminiscent of the well--known scale separation approximation, where we
assume the existence two distinct, characteristic scales of physical phenomena which we first describe
independently and later try to quantify their interaction using the perturbation expansion in the ratio of scales. The parameter $\lambda$ in their framework plays effectively 
the role of 
 the scale separation. This makes the Green--Wald formalism  an excellent tool in situations where the scale of inhomogeneities $R_\textrm{inh}$ 
is well--defined and well separated from the characteristic scale of the average, large scale metric $R_\textrm{hom}$. The larger the gap, i.e. 
the larger the ratio $R_\textrm{hom}/R_\textrm{inh}$, the better works the approximation given by perturbation theory in $\lambda$. 

Note that the multiscale foam solution does not satisfy the first  condition of applicability of the Green--Wald formalism 
because, as we have noted before in Section \ref{sec:conformalcoords}, the metric $q_N$, when taken globally, 
deviates in some places very strongly from the coarse--grained $q_0$ even for 
small $\varepsilon$. 
The solution lies thus outside the domain of validity of the Green--Wald approximation and the results of their paper do not apply to it. 
This is true despite the fact that the deviation remains small if we consider each void/overdensity pair locally, in complete
separation from all others, as if taken out from the whole hierarchy of self--similar structure. It is the collective influence of all of 
the inhomogeneities, present on all scales, which creates the large deviation of the local metric tensor from the large--scale one and invalidates the Green-Wald approach
the same way it invalidates the simple linearized GR around the flat metric. 
 
On a more fundamental level note
 that in the multiscale foam solution  \emph{there is no well--defined inhomogeneity scale}, as the inhomogeneities are present
on \emph{all} scales between $R$ and $R_\textrm{min}$. This lack of scale gap persists of course even when the ratio $R/R_\textrm{min}$
diverges to infinity, because this corresponds only to a deeper nesting level of the structure. No scale gap means that approaches based on separation of scales
do not apply.

\section{Summary and conclusions}

I have presented an exact solution of Einstein's equation with dust and no cosmological constant in which
the energy density exhibits a complicated structure  in the form of overdensities and voids extending on
many scales. The voids and overdensities form a self--similar, nested hierarchy  in which every overdensity of a given level contains the same pattern of smaller voids and 
overdensities of higher level, down to a finite, maximal nesting level. 
The solution turns out to be impossible to describe globally using the linearized gravity (first order perturbation around a flat metric) even 
if the voids and overdensities have a
very small compactness parameter $\varepsilon$, which is usually associated with very weakly relativistic, "Newtonian" solutions.

The main sign of large nonlinear GR effects is a substantial mass deficit of the solution, i.e. large difference between the ADM 
mass, i.e. the mass of the configuration measured outside, and the total mass which is the sum of the masses of all constituents. In a fully linear theory this difference vanishes.
In general relativity, for solutions without a hierarchy of nested structure, the relative difference is of the order of $\varepsilon$
 and thus negligible for objects like the Sun or Earth. It turns out however that for a carefully chosen set of 
parameters the mass deficit of the multiscale foam can be as arbitrary large no matter how small $\varepsilon$ is.

The physical mechanism behind the amplification of the mass deficit  is the accumulation of nonlinear effects of GR from many scales:  the void/overdensity pairs of 
each size give rise to  a small, positive contribution to the mass deficit. The net 
contribution from all pairs may sum up to a substantial number if the structure extends over sufficiently many scales. This mechanism 
can work for any type of nested inhomogeneities, not necessary self--similar ones I discussed in my paper.

Neither the standard first order perturbation around a flat solution nor the Green--Wald formalism can describe the foam solution in a satisfactory way. It is 
nevertheless possible to obtain a reasonably accurate value of the mass deficit if we combine the second
order perturbation theory with a coarse--graining approach: we successively smoothen out the matter distribution inhomogeneities 
level by level, calculating at each step the leading order corrections in $\varepsilon$ to the mass deficit, equal in this case to the difference of
the Newtonian binding energy before and after removing the structure.  

The mass deficit of the multiscale foam is in the leading order proportional the value of a dressed compactness parameter $\varepsilon D$, 
where $D$ is the logarithm of the ratio between the homogeneity scale and the smallest ripples scale, called the depth of structure. 
While the result has only been rigorously derived for this particular solution, it seems reasonable to assume that for 
general matter distributions with a nested multiscale structure
this holds as well if we take for $\varepsilon$ a suitably defined average value of the compactness parameter. 
The averaging should be performed over a large volume of space and 
over the overdensities of all scales down to the smallest ripples scale. 

 The results of this paper may be important for cosmology and astrophysics, especially in the connection with the backreaction problem. 
 The nested structure of large galaxy clusters, voids, filaments, etc. can potentially give rise to surprisingly large 
non--linear effects. In particular, the difference between the gravitating mass, i.e. the mass "felt" by the large scale gravitational field, and the
sum of masses of all stars, galaxies, clouds of gas and small scale structure, can be an order of magnitude larger than the na\"{i}ve expectation may suggest.
Unfortunately, it seem impossible to explain away the dark energy problem this way, as the sign of the effect is exactly opposite of the one we observe. However, 
other nonlinear effects of GR apart from mass deficit are possible. In particular,
it would be interesting to evaluate the effective pressure effects in a cosmological solution due to the presence of multi--scale structure modelled
not necessary as dust but rather as a fluid.
They would be seen as the deviation of the large--scale dynamics of the Universe from the simple FLRW behaviour.

As already noted  \cite{Ishibashi:2005sj, Green:2010qy}, the backreaction effects must be small if the metric $g$ can be written down as $g = g_0 + h$, where
 $g_0$ is the averaged metric, lacking the small--scale structure of $g$, $h$ is small everywhere and its derivatives are under control. 
 In this paper and in the previous one \cite{Korzynski:2014} I show that if these conditions are not satisfied everywhere, 
 then estimating the nonlinear relativistic effects is more difficult in general and the result depends crucially on the microscopic details of the matter distribution. 
 Namely, the clustering of compact objects on the smallest scales may give rise to large backreaction \cite{Korzynski:2014}. This is not entirely unexpected, as 
 compact binary or multi-object systems of black holes or neutron stars are obviously strongly relativistic and nonlinear. But it is less 
 intuitively clear that the presence of a
   very deep \emph{weakly relativistic} structure, extending on many scales, may cause substantial nonlinear, relativistic backreaction effects. 
   The example I have discussed illustrates that simple criteria for the value of the backreaction, based on the fact that $g$ is close to $g_0$ almost everywhere 
      or that the structure looks nonrelativistic at all scales may be quite misleading. 
  In particular, the depth of structure $D$ is an important parameter influencing the value of backreaction in the presence of multiscale hierarchy of inhomogeneities.
  Drawing the precise line between the linear and nonlinear regime in general relativity and estimating the backreaction effects in this case requires a 
  more refined approach, possibly based on the
  coarse graining approach. 
    In the next paper I will discuss a more general coarse--graining formalism for estimating various the backreaction effects, including the dynamical ones.

\section*{Acknowledgements}
The author would like to thank the Max Planck Institute for Gravitational Physics--Albert Einstein Institute in Potsdam for hospitality. The work was supported by the project \emph{
``The role of small-scale inhomogeneities in general relativity and cosmology''} (HOMING PLUS/2012-5/4), realized within the Homing Plus programme of Foundation for Polish Science, co--financed by the European Union from the Regional Development Fund. 

\appendix
\section{Matching conditions} \label{sec:appendix}

Metrics (\ref{eq:qM}) and (\ref{eq:qS}) are matched along a round sphere of area radius $r=R$. The matching conditions read
\bea
R &=& \calR\,\sin\Lambda \label{eq:match1} \\
\sqrt{1-\frac{2M}{R}} 2R &=& 2\calR \sin\Lambda \cos\Lambda \label{eq:match2}
\eea
which imply after dividing side by side 
\bea
\cos\Lambda = \sqrt{1-\frac{2M}{R}} \label{eq:cosLambda}
\eea
or (\ref{eq:sinLambda}) after a simple transformation. 
On the other hand, substituting (\ref{eq:sinLambda}) to (\ref{eq:match1}) yields (\ref{eq:calR}).

It remains to derive (\ref{eq:Mviarho}). From (\ref{eq:sinLambda}) we have $M = \frac{1}{2}R\sin^2\Lambda$, and from (\ref{eq:match1}) $\sin\Lambda = \frac{R}{\calR}$, which
taken together implies 
\bea
M = \frac{R^3}{2\calR^2}. \label{eq:MviaRcalR}
\eea We  substitute $\calR$ from (\ref{eq:rhoviacalR}) and obtain the desired result.

 In order to derive (\ref{eq:Mi}) and (\ref{eq:tildeRi}) we substitute (\ref{eq:match1}) to (\ref{eq:MviaRcalR}) to obtain 
 \bea
 M = \frac{\calR\sin^3\Lambda}{2}. \nonumber
 \eea
Since both exterior and interior Schwarzschild solutions are matched to the same $S^3$ solution, this equation must also hold with $M_i$ and $\Lambda_i$
substituted for 
 $M$ and $\Lambda$, hence (\ref{eq:Mi}). The same reasoning applied to (\ref{eq:match1}) yields (\ref{eq:tildeRi}).

\section*{References}
\bibliographystyle{iopart-num}
\bibliography{structure}

\end{document}